\documentclass[a4paper,11pt]{article}

\usepackage{jheppub}

\usepackage{tensor}
\usepackage[utf8]{inputenc}
\usepackage{slashed}
\usepackage{float}
\usepackage{amsmath,amssymb}
\usepackage{dsfont}
\usepackage{hyperref}
\usepackage{graphicx}
\usepackage{enumitem}
\usepackage{mathtools}
\usepackage{bbold}
\usepackage{multirow}
\usepackage{ytableau}
\usepackage{youngtab}
\usepackage{braket}
\usepackage{soul}
\usepackage{cancel}
\usepackage{xcolor}
\usepackage{physics}
\usepackage{bm}
\usepackage[normalem]{ulem}
\usepackage{subcaption}
\usepackage{booktabs}

\usepackage{lipsum}

\usepackage{bbold}
\usepackage{tikz}
\allowdisplaybreaks

\definecolor{nicered}{rgb}{0.7,0.1,0.1}
\definecolor{nicegreen}{rgb}{0.1,0.5,0.1}
\definecolor{violet}{rgb}{0.7,0.3,0.3}
\definecolor{rhodamine}{rgb}{0.9,0,0.58}
\hypersetup{colorlinks,citecolor= nicegreen,linkcolor= nicered}

\newcommand{\mc}[1]{\mathcal{#1}}
\newcommand{\mrm}[1]{\mathrm{#1}}

\newcommand{\re}{\mathrm{Re}}
\newcommand{\im}{\mathrm{Im}}

\newcommand{\rre}{\text{Re}}
\newcommand{\iim}{\text{Im}}

\title{Current and future constraints on heavy New Physics from $\tau$ weak dipole moments}

\author[a, b]{Nejc Košnik,}
\emailAdd{nejc.kosnik@ijs.si}
\affiliation[a]{Jo\v zef Stefan Institute, Jamova 39, 1000  Ljubljana, Slovenia}
\affiliation[b]{%
  Faculty of Mathematics and Physics, University of Ljubljana, Jadranska 19, 1000 Ljubljana,
Slovenia
}%
\author[a]{Zachary Polonsky,}
\emailAdd{zach.polonsky@ijs.si}

\author[a]{Aleks Smolkovič}
\emailAdd{aleks.smolkovic@ijs.si}

\begin{document}


\abstract{
We study the weak magnetic and electric dipole moments of the $\tau$ lepton as precision tests of the Standard Model (SM) and probes of heavy New Physics (NP). We present an updated SM prediction for the $\tau$ weak magnetic dipole moment at one loop, including a careful assessment of theoretical uncertainties from electroweak scheme dependence. Working within the SM Effective Field Theory, we derive comprehensive current constraints on the $\tau$ dipole operators from a combination of observables: the $\tau$ weak and electromagnetic dipole moments, high-mass Drell-Yan tails at the LHC, $Z$ partial decay widths, and the electron electric dipole moment. Finally, we assess the prospects of measuring the SM value of the $\tau$ weak magnetic moment at the FCC-$ee$ Tera-$Z$ run, and project the sensitivities of the leading observables to heavy NP at FCC-$ee$ and HL-LHC, incorporating informed assumptions about improvements of systematic uncertainties. We find that the $\tau$ weak dipole moments are already among the leading probes of the $\tau$ dipole operators, and will become increasingly dominant at future colliders.
}

\maketitle

\section{Introduction}

The dipole moments of charged leptons are among the most powerful low-energy probes of fundamental interactions, offering sensitivity to New Physics (NP) at scales far beyond those directly accessible at colliders. For the electron and muon, the anomalous magnetic moments, $(g-2)_{e,\mu}$, have been measured with extraordinary precision via spin precession in storage rings~\cite{VanDyck:1987ay,Hanneke_2008,Fan:2022eto,ParticleDataGroup:2026aaa,Muong-2:2025xyk}, a technique not feasible for the $\tau$ due to its short lifetime. This same lifetime, on the other hand, is an advantage for the measurement of weak dipole moments, which arise from couplings to the $Z$ and can only be accessed experimentally through polarized-$\tau$ observables~\cite{Bernabeu:1993er,Bernabeu:1994wh,ALEPH:2001uca,ALEPH:2002kbp}. Since the $\tau$ decays within the detector volume before any depolarization occurs, its spin state is directly accessible from the angular distributions of its decay products~\cite{Bernabeu:1993er,Bernabeu:1994wh,Bernabeu:2007rr,Bernabeu:2008ii,Gogniat:2025eom,Crivellin:2021spu,Hoferichter:2025ijh,Hoferichter:2025zjp,Eidelman:2007sb,Du:2026yoi}. As a consequence, the $\tau$ weak dipole moments are arguably the only ones that are feasibly measurable for any charged lepton, motivating a precise theoretical understanding and experimental program for these observables.

The Standard Model (SM) prediction for the $\tau$ weak magnetic dipole moment was computed in Ref.~\cite{Bernabeu:1994wh}, yielding $\mu_\tau^w = -(2.10 + 0.61i)\times 10^{-6}$, where the imaginary part arises from absorptive parts of loop diagrams. The $C\!P$-violating weak electric dipole moment $d_\tau^w$ arises at four loops~\cite{Pospelov:2013sca,Yamaguchi:2020dsy,Yamaguchi:2020eub} and is predicted to be negligibly small in the SM. Together, these predictions set a clear experimental target: for $\mu_\tau^w$, sensitivity at the level of $\mathcal{O}(10^{-6})$ is required before the SM prediction can be meaningfully probed, while for $d_\tau^w$ any nonzero experimental signal would directly indicate NP. The most stringent experimental constraints on both $\mu_\tau^w$ and $d_\tau^w$ currently come from the ALEPH experiment at LEP~\cite{ALEPH:2002kbp}, which achieved a sensitivity of $\mathcal{O}(10^{-3})$ from $e^+e^- \to \tau^+\tau^-$ production at the $Z$-pole  (for phenomenological discussions, see e.g. Refs.~\cite{Gonzalez-Sprinberg:2000lzf,Huang:1998wh,Escribano:1996wp}). The Tera-$Z$ phase of the Future Circular Collider (FCC-$ee$) is expected to produce $\mathcal{O}(10^{12})$ $Z$ bosons~\cite{FCC:2018byv}, offering a dramatic reduction in statistical uncertainties, and a significant expected improvement in probing the $\tau$ weak dipole moments~\cite{Buttazzo:2026amk}. However, at this level of luminosity, systematic uncertainties will become the dominant bottleneck in polarized-$\tau$ observables~\cite{FCC:2018byv}, and as such special care must be taken when projecting the current bounds. The potential reach of the FCC-$ee$ makes an updated calculation of the SM prediction for $\mu_\tau^w$ including a careful assessment of theory uncertainties both timely and warranted.

Assuming only SM light degrees of freedom, linearly realized electroweak symmetry breaking, and a sufficient scale separation between the electroweak scale and the scale of NP, the effects of potential new heavy degrees of freedom can be efficiently encoded in the Wilson coefficients of the SM Effective Field Theory (SMEFT)~\cite{Buchmuller:1985jz,Grzadkowski:2010es,Brivio:2017vri,Isidori:2023pyp}. In a model-independent, bottom-up approach, these act as an agnostic description of NP effects in observables, the scales of which can, through the inclusion of Renormalization Group (RG) evolution, be connected with a heavy NP scale~\cite{Jenkins:2013zja,Jenkins:2013wua,Alonso:2013hga}. Through the process of matching, these results can eventually be reinterpreted in the context of generic UV models~\cite{deBlas:2017xtg, Carmona:2021xtq,Fuentes-Martin:2022jrf}. 

In this framework, the $\tau$ weak dipole moments receive contributions from the $\tau$ dipole operators at dimension six, and interpreting any measurement in terms of NP requires a global analysis: the same operators are constrained by a rich set of complementary observables, including the $\tau$ electromagnetic dipole moments, high-mass Drell-Yan tails at the LHC, $Z$ partial decay widths, and, through higher-order effects, the electric dipole moment of the electron. In this work, we perform a comprehensive SMEFT study of the $\tau$ dipole operators, incorporating all relevant complementary constraints. In addition, we provide projected bounds on the relevant Wilson coefficients expected from FCC-$ee$ and the High-Luminosity LHC (HL-LHC), offering a detailed picture of the NP sensitivity achievable at future experiments, with assumptions about systematic uncertainties informed by analyses of similar processes by the FCC collaboration~\cite{FCC:2018byv}.

This paper is organized as follows. In Section~\ref{sec:weak}, we introduce the weak dipole moments of the $\tau$, present an updated SM prediction for $\mu_\tau^w$, and review the current experimental status. Section~\ref{sec:NP} is devoted to the SMEFT analysis of heavy new physics: after setting up the EFT framework, we discuss the current sensitivities of the complementary observables and the implications for UV models. Future prospects for both testing the SM prediction and probing heavy NP are discussed in Section~\ref{sec:future}. We conclude in Section~\ref{sec:conc}. Detailed analytic results for the SM prediction of the $\tau$ weak magnetic moment are collected in Appendix~\ref{app:analyticRes}.

\section{Weak multipole moments of the $\tau$}
\label{sec:weak}
Multipole moments are typically computed by considering matrix elements with single-current insertions between initial and final states. In the case of the insertion of the weak neutral current, $j_w^\mu$, between the vacuum and a $\tau^+\tau^-$ final state, such a matrix element can be decomposed into a complete set of Lorentz structures with corresponding form factors~\cite{Hollik:1997vb}
\begin{equation}\begin{split}\label{eq:tauMultipoleME}
    \mel{\tau^+(p_1)\tau^-(p_2)}{j_w^\mu}{0} = e\bar{u}(p_2)\Big[&\gamma^\mu F^w_1(q^2) + \frac{i\sigma^{\mu\nu}q_\nu}{2m_\tau} F^w_2(q^2) + \frac{\sigma^{\mu\nu}\gamma_5 q_\nu}{2m_\tau} F^w_3(q^2)\\
    &+ \gamma^\mu \gamma_5 F^w_4(q^2) - q^\mu\big(F^w_S(q^2) + F^w_P(q^2)\gamma_5\big)\Big]v(p_1) \,,
\end{split}\end{equation}
where the weak neutral current for $\tau$ is defined as\footnote{When considering the matrix element in Eq.~\eqref{eq:tauMultipoleME} beyond tree-level in the electroweak interaction, one must also consider couplings of the weak neutral current to other fermions as well as electroweak bosons and the Higgs. See e.g. Ref.~\cite{Denner:1991kt} for details.}
\begin{equation}
    j_w^\mu =   e\,\bar \tau \gamma^\mu ( g_L P_L + g_R P_R) \tau \,,
\end{equation}
with $P_{L(R)}=\frac{1}{2}(1\mp\gamma_5)$.
Here the right- and left-handed couplings of the $Z$ to the charged leptons are given by
\begin{equation}\label{eq:gRgL}
    g_R = \frac{s_w}{c_w}\,, \qquad g_L = \frac{2 s_w^2 - 1}{2s_w c_w}\,,
\end{equation}
and $s_w$ $(c_w)$ is the sine (cosine) of the weak mixing angle. The total momentum of the $\tau^+\tau^-$ final state is denoted by $q^\mu = p_1^\mu + p_2^\mu$\footnote{This matrix element is often given in terms of one incoming and one outgoing fermion where $q$ represents the momentum transfer between the initial and final states. However, the notation of Eq.~\eqref{eq:tauMultipoleME} is better adapted to $Z$-pole observables and is consistent with that used in Ref.~\cite{ALEPH:2002kbp}.}, $e$ is the positron charge, and $\sigma^{\mu\nu}=\frac{i}{2}[\gamma^\mu,\gamma^\nu]$. For an on-shell $Z$-boson, $q^2=M_Z^2$, the multipole moments are defined in terms of the form factors as (following the notation of Ref.~\cite{ALEPH:2002kbp})
\begin{equation}\label{eq:multipoles}
    v_\tau^w = F^w_1(M_Z^2)\,, \quad
    a_\tau^w = - F^w_4(M_Z^2)\,, \quad
    \mu_\tau^w = F^w_2(M_Z^2)\,, \quad
    d_\tau^w = F^w_3(M_Z^2) \,,
\end{equation}
where $v_\tau^w$ and $a_\tau^w$ represent the vector and axial-vector couplings to the $Z$ including quantum corrections, while $\mu_\tau^w$ and $d_\tau^w$ give the weak magnetic (wMDM) and weak electric (wEDM) dipole moments, respectively. Additionally, the last term in Eq.~\eqref{eq:tauMultipoleME} proportional to $q^\mu$ vanishes when the current is coupled to an on-shell $Z$-boson.

The weak multipoles in Eq.~\eqref{eq:multipoles} are analogous to the corresponding standard electromagnetic (EM) multipoles, but with several key differences. First, since they are defined at $q^2=M_Z^2$, they are directly probed by lepton colliders tuned to the $Z$-pole such as LEP or the Tera-$Z$ run at FCC-$ee$. This is in contrast to the EM form factors of the $\tau$, which are typically probed at colliders via off-shell photon exchanges, and must be extrapolated to $q^2=0$ to obtain the true values of the moments~\cite{Bernabeu:2007rr,Bernabeu:2008ii,Gogniat:2025eom,Gogniat:2026zvf,Hoferichter:2025ijh,Hoferichter:2025zjp,Biebel199753,Gau:1997cn,Eidelman:2016aih,Cornet:1995pw,Dyndal:2020yen,Atag:2010ja,Du:2026yoi}. Second, due to the parity violation of the weak interaction, $v_\tau^w$ and $a_\tau^w$ appear at tree-level with similar magnitudes, $|v_\tau^w|\sim |a_\tau^w| \sim 1$, whereas the EM axial-vector form-factor only arises from higher-order electroweak corrections. This opens new channels for interference between the weak dipoles and $a_\tau^w$, presenting additional observables where these moments could be observed.

On the other hand, the wMDM and wEDM are highly suppressed in the SM: the former arises at one-loop in the electroweak couplings, giving $\mu_\tau^w\sim 10^{-6}$~\cite{Bernabeu:1993er,Bernabeu:1994wh}, while the latter is $C\!P$-violating, and only arises at higher loop orders, further suppressed by tiny neutrino masses or the GIM mechanism in the quark sector~\cite{Pospelov:2013sca,Yamaguchi:2020dsy,Yamaguchi:2020eub}. These strong suppressions in the SM make the weak dipole moments potentially powerful probes of NP. 

\subsection{Updated Standard Model predictions}\label{sec:smRes}
\begin{figure}[t]
 \centering
 \includegraphics[width=0.99\textwidth]{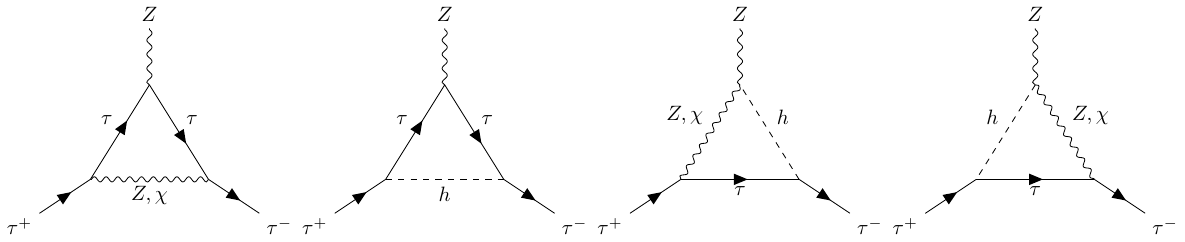}\\[0.5em]
 \includegraphics[width=0.5\textwidth]{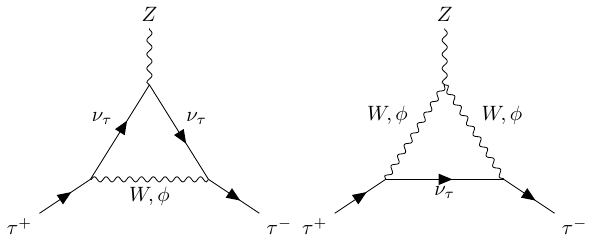}\\[0.5em]
 \includegraphics[width=0.25\textwidth]{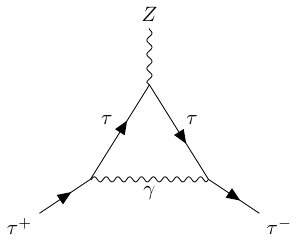}
 \caption{Three independent sets of gauge-invariant Feynman diagrams involving internal $Z$-bosons, neutral Goldstones, and Higgs bosons (top), $W$-bosons and charged Goldstones (middle), and photons (bottom). Wavy internal lines represent both massive vectors and corresponding Goldstone bosons if applicable.}
 \label{fig:dipolefeyndias}
\end{figure}

The $\tau$ wMDM was first computed in Ref.~\cite{Bernabeu:1994wh} with the quoted value
\begin{equation}\label{eq:bernabeuRes}
    \mu_\tau^w = -(2.10 + 0.61i)\times 10^{-6}
\end{equation}
and no given uncertainty. As we will discuss in Sec.~\ref{sec:measureSMDipole}, depending on the assumptions for the improvement of systematic uncertainties, future measurements of $\mu_\tau^w$ at FCC-$ee$ can potentially reach $\mathcal{O}(10^{-6})$ precision, and so we present in this section an updated value of the SM prediction with a full theoretical uncertainty analysis. Additionally, as an explicit cross-check, we independently re-computed the analytic results of Ref.~\cite{Bernabeu:1994wh} in a general $R_\xi$ gauge, expanding to linear order in the gauge parameters around Feynman gauge, i.e. expanding in $\Xi_i = (1 - \xi_i)$ for $i=\gamma,Z,W$, and verifying that each gauge-invariant set of diagrams shown in Fig.~\ref{fig:dipolefeyndias} is independent of $\Xi_i$\footnote{We also verified the finiteness of $\mu_\tau^w$ and $d_\tau^w=0$ as additional explicit checks of the calculation.}. The calculation was performed using a combination of the \texttt{MaRTIn} program~\cite{Brod:2024zaz} and custom \texttt{FORM}~\cite{Vermaseren:2000nd} code to reduce tensor integrals and \texttt{Package-X}~\cite{Patel:2016fam} to evaluate the resulting scalar integrals. Analytic results are given in Appendix~\ref{app:analyticRes}.

The leading contribution to $\mu_\tau^w$ is dependent on the choice of electroweak scheme, as well as renormalization scale. It is well-known (see e.g.~Refs.~\cite{Biekotter:2025nln,Brod:2010hi,Brod:2021qvc,Bobeth:2013tba}) that this leading-order scheme dependence can introduce additional theoretical uncertainties which can be substantially larger than naive estimates of NLO EW corrections, and can only be reduced by computing at higher orders in electroweak couplings\footnote{For example, in Ref.~\cite{Bobeth:2013tba}, it was shown that choosing $s_w^2=0.2231$ in the on-shell scheme versus $s_w^2=0.2314$ in the $\overline{\text{MS}}$ scheme leads to a $\sim7\%$ uncertainty in the $B_s\to \mu^+\mu^-$ branching ratio, as opposed to the naive expectation for NLO EW corrections $|g_2^2/(16\pi^2)\log(m_b^2/M_Z^2)|\sim 1-2\%$.}. In order to estimate the uncertainty arising from the choice of scheme, we compute $\mu_\tau^w$ in five different schemes with input parameters given in Tab.~\ref{tab:schemeinputs}.

\begin{table}[t]
    \centering
    \caption{Input schemes used to evaluate the scheme uncertainty in $\mu_\tau^w$. All masses are assumed to be in the on-shell scheme, $G_F$ corresponds to the all-orders Wilson coefficient of muon decay, and $\alpha(M_Z)$ and $s_w^2(M_Z)$ are assumed to be in the $\overline{\text{MS}}$ scheme. Naming convention is chosen to be consistent with Ref.~\cite{Biekotter:2025nln}.}
    \label{tab:schemeinputs}
    \begin{tabular}{ll}
        \toprule
        Scheme & Input Parameters \\
        \midrule
         $v_\mu$ & $G_F$, $s_w^2(M_Z)$, $M_Z$, $m_\tau$, $M_h$ \\
         $v_\alpha$ & $\alpha(M_Z)$, $s_w^2(M_Z)$, $M_Z$, $m_\tau$, $M_h$ \\
         $\alpha_\mu$ & $G_F$, $M_W$, $M_Z$, $m_\tau$, $M_h$ \\
         $\alpha$ & $\alpha(M_Z)$, $M_W$, $M_Z$, $m_\tau$, $M_h$\\
         LEP & $G_F$, $\alpha(M_Z)$, $M_Z$, $m_\tau$, $M_h$ \\
         \bottomrule
    \end{tabular}
\end{table}

\begin{table}[t]
    \centering
    \caption{Input parameters used for the computation of the $\tau$ wMDM. All values taken from the PDG~\cite{ParticleDataGroup:2026aaa} where available. To determine the input of $\alpha(M_Z)^{-1}$, we use the methodology outlined in the PDG (see also Refs.~\cite{Fanchiotti:1992tu,Erler:1998sy,Erler:2017knj,Davier:2019can,Keshavarzi:2019abf,Chetyrkin:1996cf}) with $\Delta\alpha_{\text{lep}}=0.031497(2)$~\cite{Sturm:2013uka} and $\Delta\alpha_{\text{had}}(M_Z)=0.02775(15)$, the latter of which accommodates determinations from both lattice and $e^+e^-\to\text{hadrons}$ data~\cite{Cirigliano:2026ios,Davier:2019can,Keshavarzi:2019abf,Ce:2022eix,Erler:2023hyi,Conigli:2025qvh}.}
    \label{tab:paraminputs}
    \begin{tabular}{ll}
        \toprule
        Parameter & Value \\
        \midrule
         $M_Z$ & $91.1880(20)$ GeV \\
         $M_W$ & $80.3692(133)$ GeV \\
         $M_h$ & $125.20(11)$ GeV \\
         $m_\tau$ & $1.77693(9)$ GeV \\
         $\alpha(M_Z)^{-1}$ & $127.941(21)$\\
         $s_w^2(M_Z)$ & $0.23121(4)$ \\
         $G_F$ & $1.1663785(6)\times 10^{-5}$ GeV$^{-2}$ \\
         \bottomrule
    \end{tabular}
\end{table}

Necessary parameters which are not given as inputs are determined through the following tree-level relations\footnote{It should be noted that these relations receive corrections at the loop-level when converting between schemes~\cite{Brod:2021qvc,Jegerlehner:2001fb,Jegerlehner:2002em}, but these expressions suffice to the level of precision that we consider.}
\begin{equation}
    G_F = \frac{\pi\alpha}{\sqrt{2}M_W^2s_w^2}\,, \quad
    c_w = \sqrt{1 - s_w^2}\,, \quad
    M_W = c_w M_Z \,.
\end{equation}
Parametric inputs are shown in Tab.~\ref{tab:paraminputs}.
Note that the value of $\alpha$ listed in the table is defined in the five-flavor theory, and $s_w^2(M_Z)$ is the $\overline{\text{MS}}$ value in the scheme with $\alpha\log(m_t/M_Z)$ terms decoupled in $\gamma-Z$ mixing~\cite{ParticleDataGroup:2026aaa}. In our calculation, we do not include any top-quark re-coupling corrections. These effects are formally higher-order than the present computation, and are expected to be significantly smaller than the allocated theory uncertainty, so we neglect them.

\begin{table}[t!]
\centering
\caption{Real and imaginary values of the $\tau$ wMDM in the five input schemes defined in Table~\ref{tab:schemeinputs}. In all cases, the first uncertainty corresponds to the uncertainty arising from varying the $\overline{\text{MS}}$ scale in the range $\mu\in[60, 320]$ GeV, while the second corresponds to parametric uncertainty.}
\label{tab:schemeResults}
\begin{tabular}{lll}
    \toprule
    Scheme & $\rre(\mu_\tau^w)\times 10^6$ & $\iim(\mu_\tau^w)\times 10^6$ \\
    \midrule
    $v_\mu$ & $-2.0474(586)(2)$ & $-0.6118(249)(1)$ \\
    $v_\alpha$ & $-2.0621(410)(4)$ & $-0.6162(301)(1)$ \\
    $\alpha_\mu$ & $-1.9375(0)(35)$ & $-0.6592(0)(15)$ \\
    $\alpha$ & $-2.0005(75)(20)$ & $-0.6806(26)(21)$ \\
    LEP & $-2.0811(181)(9)$ & $-0.5975(77)(4)$ \\
    \bottomrule
\end{tabular}

\end{table}

Numerical results in the five different input schemes are given in Tab.~\ref{tab:schemeResults}, and can be highly incompatible for the given uncertainties, signaling that the uncertainty estimate is being underestimated in some, if not all, schemes. This is particularly apparent in the $\alpha$ and $\alpha_\mu$ schemes, which feature significantly smaller scale variance due to the treatment of $s_w^2$ in the on-shell scheme, $s_w^2=1-(M_W^2/M_Z^2)^{\text{pole}}$. However, it has been found that on-shell $s_w^2$ receives large contributions from the top-quark~\cite{Jegerlehner:2001fb,Jegerlehner:2002em}, and poor perturbative convergence has been observed in several cases when using these schemes once higher-order electroweak corrections are computed, see e.g.~Refs.~\cite{Bobeth:2013tba,Brod:2021qvc} for examples in $B_s\to\mu^+\mu^-$ and neutral meson mixing. With this in mind, in order to quote a single SM prediction, we neglect the $\alpha$ and $\alpha_\mu$ schemes and take a weighted average of the remaining schemes (which are consistent within $1\sigma$), inflating the resulting uncertainty by the largest scale uncertainty in Table~\ref{tab:schemeResults} to account for a potentially large scheme-dependence. We find
\begin{equation}\begin{split}\label{eq:F2result}
    &\rre\left(\mu_\tau^w\right) = - 2.0757(586)_{\text{scheme}}(160)_{\text{scale}}(6)_{\text{param}}\times 10^{-6}\,,\\[0.5em]
    &\iim\left(\mu_\tau^w\right) = - 0.5997(586)_{\text{scheme}}(71)_{\text{scale}}(3)_{\text{param}}\times 10^{-6}\,,
\end{split}\end{equation}
where the uncertainties reflect the inflated uncertainty to account for scheme-dependence, scale variation after the weighted average, and parametric uncertainty, respectively. Here, the scheme uncertainty should be viewed as a systematic uncertainty, effectively specifying a range of equivalent SM predictions.\footnote{This method of inflating uncertainties results in a value which is consistent with all schemes within (or just over, in the case of the $\alpha$ scheme) $1\sigma$, aside from the $\alpha_\mu$ scheme whose uncertainties are likely highly underestimated. That said, a slightly more conservative choice of the SM prediction enforcing consistency with all schemes can be given by the flat ranges
    ${\rre\left(\mu_\tau^w\right) \in [-2.11,-1.93]\times10^{-6}}$,
    ${\iim\left(\mu_\tau^w\right) \in [-0.66,-0.59]\times 10^{-6}}$.
}

Eqs.~\eqref{eq:F2result} are in good agreement with the results from Ref.~\cite{Bernabeu:1994wh}. Although the scheme uncertainty can only be reduced with a full electroweak two-loop evaluation of $\mu_\tau^w$, this should not be necessary since future experiments will be unlikely to measure the SM value of the $\tau$ wMDM to 10\% precision.

\subsection{Experimental status}\label{sec:expStatus}
To date, the ALEPH collaboration has set the best bounds on the weak dipole moments of the $\tau$ using an integrated luminosity of $155\text{ pb}^{-1}$ of $e^+ e^-$ collisions near the $Z$-pole at LEP~\cite{ALEPH:2002kbp}. The results read
\begin{equation}\begin{split}\label{eq:alephBounds}
    \re\big(\mu_\tau^w\big) = (-0.33\pm0.49)\times 10^{-3}\,&, \quad
    \im\big(\mu_\tau^w\big) = (-0.99\pm1.01)\times 10^{-3}\,, \\[0.5em]
    \re\big(d_\tau^w\big) = (-0.11\pm0.45)\times 10^{-3}\,&, \quad
    \im\big(d_\tau^w\big) = (-0.08\pm1.02)\times 10^{-3}\,,
\end{split}\end{equation}
where the combined statistical and systematic errors are written, and with small correlations as given in Ref.~\cite{ALEPH:2002kbp}. These were found using an angular analysis of $e^+e^-\to \tau^+\tau^-$ for polarized $\tau^\pm$ final states which probe the Lorentz structure of the $Z\tau\tau$-vertex. The differential cross-section is given by
\begin{equation}\label{eq:crossSecDecomp}
    \frac{d\sigma}{d\cos\theta_\tau}\big(\vec s_+,\vec s_-\big) = R_{00} + \sum_{\mu = 1}^3 R_{\mu0}s_+^\mu + \sum_{\nu = 1}^3 R_{0\nu}s_-^\nu + \sum_{\mu,\nu=1}^3 R_{\mu\nu}s_+^\mu s_-^\nu\,,
\end{equation}
where $s^\mu_\pm$ and $\vec s_{\pm}$ are the spin four-vectors and spin quantization axes of the $\tau^{\pm}$, respectively, and $\theta_\tau$ is the production angle of the $\tau$. The spatial coordinate axes are chosen in Ref.~\cite{ALEPH:2002kbp} such that the $z$ ($3$) axis is along the three-momentum of the $\tau^+$, and the $y-z$ ($2-3$) plane includes the three-momentum of the incoming $e^+$.

In hadronic $\tau$ decays, the spin of the $\tau^\pm$ is correlated with the final-state momentum of the hadron, $h$. In the rest frame of the decaying $\tau$, this can be given in terms of the polarization analyzer~\cite{ALEPH:2002kbp,Bernabeu:1993er,JADACH1991275,JADACH1993361} (see also e.g.~Ref.~\cite{Gogniat:2025eom} for a recent discussion)
\begin{equation}
    \frac{d\Gamma(\tau^\pm(\vec s_\pm) \to h^\pm(\vec p)\,\nu_\tau)}{\Gamma(\tau^\pm\to h^\pm \nu_\tau)} = \frac{1}{4\pi}\big(1 + \alpha_h\,\hat p\cdot\vec s_{\pm}\big)d\Omega_{h^\pm} \,,
\end{equation}
where $\vec p$ is the three-momentum of the final-state hadron, $h$, and $\alpha_h$ is the polarization analyzer which depends on the spin of $h$. For example, when summing over all spin components of $h$, $\alpha_h$ is given by~\cite{Gogniat:2025eom}
\begin{equation}
    \alpha_h = \left\{
    \begin{array}{ccc}
        1\,, && j_h = 0\\
        \frac{m_\tau^2 - 2m_h^2}{m_\tau^2 + 2m_h^2}\,, && j_h = 1 
    \end{array}\right.\,.
\end{equation}
The polarization analyzer can be decomposed further for the case of $j_h=1$ into transverse and longitudinal components.

The different terms in Eq.~\eqref{eq:crossSecDecomp} admit different angular dependencies on the four multipole moments in Eq.~\eqref{eq:multipoles}. For example, the longitudinally-polarized components are probed through the linear combination
\begin{equation}
    \frac{R_{03} + R_{30}}{R_{00}} \equiv \mathcal{P}_\tau(\cos\theta_\tau) = -\frac{\mathcal{A}_\tau(1 + \cos^2\theta_\tau) + 2\mathcal{A}_e\cos\theta_\tau}{(1 + \cos^2\theta_\tau) + 2\mathcal{A}_\tau\mathcal{A}_e\cos\theta_\tau}\,,
\end{equation}
where
\begin{equation}
    \mathcal{A}_\ell = \frac{2 \,\re\big(a_\ell^w{v_\ell^w}^*\big)}{|a_\ell^w|^2 + |v_\ell^w|^2}\,.
\end{equation}
Performing an angular average of $\mathcal{P}_\tau(\cos\theta_\tau)$ extracts $\mathcal{A}_\tau$, which serves as a precision test of the electroweak couplings~\cite{ALEPH:2005ab,ALEPH:2001uca}.

The weak dipoles are accessible through measurements of the transverse polarizations\footnote{In this context transverse polarization direction is in the
$e^+ \tau^+$ plane and perpendicular to the $\tau^+$ momentum.} of the final-state $\tau$. The most sensitive terms in Eq.~\eqref{eq:crossSecDecomp} to the real parts of the wMDM and wEDM are~\cite{ALEPH:2002kbp}
\begin{equation}\begin{split}\label{eq:reWeakEDMTerms}
    R_{02} + R_{20} &\propto \frac{2}{\gamma_\tau}\sin\theta_\tau|v^w_\tau|^2\re\big(v^w_e {a_e^w}^*\big) + \frac{1}{\gamma_\tau}\sin\theta_\tau\cos\theta_\tau\big(|a_e^w|^2 + |v_e^w|^2\big)\re\big(v_\tau^w {a_\tau^w}^*\big) \\[0.5em]
    &+ \gamma_\tau\sin\theta_\tau\cos\theta_\tau\big(|a_e^w|^2 + |v_e^w|^2\big)\re\big(a_\tau^w {\mu_\tau^w}^*\big) + 2\gamma_\tau\sin \theta_\tau \re\big(v_e^w {a_e^w}^*\big)\re\big(v_\tau^w{\mu_\tau^w}^*\big)\,, \\[1em]
    R_{01} - R_{10} &\propto  - \gamma_\tau\sin\theta_\tau\cos\theta_\tau\big(|a_e^w|^2 + |v_e^w|^2\big)\re\big(a_\tau^w {d_\tau^w}^*\big) - 2\gamma_\tau\sin\theta_\tau\re\big(v_e^w {a_e^w}^*\big)\re\big(v_\tau^w{d_\tau^w}^*\big)\,,
\end{split}\end{equation}
respectively, where $\gamma_\tau = M_Z/2m_\tau$ for $e^+e^-\to \tau^+\tau^-$ at the $Z$-pole is the boost factor of the final-state $\tau^\pm$. Note that in Eq.~\eqref{eq:reWeakEDMTerms}, we have kept only terms linear in the weak dipole moments which are enhanced by $\gamma_\tau$.

The fact that $a_\ell^w$ is generated at tree-level in the SM is clearly advantageous for searches for the weak dipole moments of the $\tau$, since all cross-terms in Eq.~\eqref{eq:reWeakEDMTerms} are at least linear in $a_e^w$ or $a_\tau^w$, making measurements of $\mu_\tau^w$ and $d_\tau^w$ feasible just from polarized $\tau$ data on the $Z$-pole. This is in contrast to measurements of the $\tau$ EM dipoles which must rely on polarized electron beams~\cite{Bernabeu:2007rr,Bernabeu:2008ii,Gogniat:2025eom,Gogniat:2026zvf,Hoferichter:2025ijh,Hoferichter:2025zjp}, radiative processes such as $e^+e^-\to \tau^+\tau^-\gamma$~\cite{Biebel199753,Gau:1997cn,L3:1998gov} and $\tau\to\ell \nu\bar\nu \gamma$~\cite{Eidelman:2016aih}, or photon-photon collisions~\cite{Cornet:1995pw,Dyndal:2020yen,Atag:2010ja,CMS:2024qjo,ATLAS:2022ryk,DELPHI:2003nah}.

\section{Constraints on heavy New Physics}
\label{sec:NP}
In this section, we derive constraints on heavy NP effects in the $\tau$ weak dipole moments in a model-independent way. We set up the SMEFT framework, identify the relevant dimension-six operators and discuss their RG evolution and mixing, before turning to the complementary observables that constrain the same operators. The combined phenomenological analysis and its implications for UV models are presented subsequently.

\subsection{Effective Field Theory setup}\label{sec:eftSetup}

We use the following definition of the SMEFT Lagrangian
\begin{equation}
\mathcal{L_{\mathrm{SMEFT}}} = \mathcal{L_{\mathrm{SM}}} + \sum_{\mathcal{O}_i=\mathcal{O}_i^\dagger} {C_i} \mathcal{O}_i + \sum_{\mathcal{O}_i\neq \mathcal{O}_i^\dagger} \left( {C_i}\mathcal{O}_i + {C_i^\ast} \mathcal{O}_i^\dagger \right) \,,
\label{eq:SMEFT}
\end{equation}
where $C_i$ are Wilson coefficients (WCs) and $\mathcal{O}$ are hermitian or non-hermitian local operators. In this work we consider mass dimension $6$ operators in the Warsaw basis~\cite{Grzadkowski:2010es}, the exact forms of which we define in the subsequent discussion when necessary.

First, we focus on the tree-level effects of the lepton dipole operators, defined as
\begin{equation}
\begin{split}
\mathcal{O}_{eW}^{pr} &= (\bar{l}_p \sigma^{\mu\nu} \tau^I e_r) H W_{\mu\nu}^I \,,\\
\mathcal{O}_{eB}^{pr} &= (\bar{l}_p \sigma^{\mu\nu} e_r) H B_{\mu\nu}\,,
\end{split}
\end{equation}
where $l_p$ and $e_r$ are the left-handed lepton doublets and right-handed lepton singlets with flavor indices $p$ and $r$, respectively, $H$ is the Higgs doublet, and $W_{\mu\nu}$ and $B_{\mu\nu}$ are the $SU(2)_L$ and $U(1)_Y$ field strengths. After EWSB, these operators rotate into the following symmetry-broken operators, written in the lepton mass-basis:
\begin{equation}
\begin{split}
    \mathcal{O}_{e \gamma}^{pr}&= \bar{e}_L^p \sigma^{\mu\nu} e_R^r F_{\mu\nu} \,,\\
    \mathcal{O}_{e Z}^{pr} &= \bar{e}_L^p \sigma^{\mu\nu} e_R^r Z_{\mu\nu} \,,\\
\end{split}
\end{equation}
via the relations
\begin{equation}
\label{eq:CegammaCeZ}
\begin{split}
    C_{e\gamma}^{pr} &= \frac{v}{\sqrt{2}} \left[ -C_{eW}^{pr} s_w + C_{eB}^{pr} c_w\right] \,,\\
    C_{eZ}^{pr} &= -\frac{v}{\sqrt{2}} \left[C_{eW}^{pr} c_w + C_{eB}^{pr} s_w\right] \,,\\
\end{split}
\end{equation}
where $v$ is the Higgs vacuum expectation value. Mapping to the conventions defined in Eqs.~\eqref{eq:tauMultipoleME}, \eqref{eq:multipoles}, we obtain the following contributions to the weak magnetic dipole and electric dipole moments\footnote{We are only concerned with constraints on heavy NP, which will only give leading contributions to the real parts of the weak dipole moments. The imaginary parts are generated from absorptive parts of loops containing light degrees of freedom which can go on-shell, and are potentially sensitive to light NP.}
\begin{equation}
\label{eq:wdipolesSMEFT}
\begin{split}
    \re \big(\mu^w_\ell\big) &= -\frac{4 m_\ell}{e} \re \big[C_{eZ}^{\ell\ell}\big] \,,\\
    \re \big(d^w_\ell\big) & = \frac{4 m_\ell}{e} \im \big[C_{eZ}^{\ell\ell}\big]\,,
\end{split}
\end{equation}
where $\ell$ denotes the lepton flavor. As in this paper we are most interested in the $\tau$ weak dipole moments, we will focus on $\mathcal{O}_{eW}$ and $\mathcal{O}_{eB}$ with only the $\tau$ indices active at the NP scale $\Lambda$. In this case e.g.~$\re (\mu^w_\tau) \propto \re [C_{eZ}^{33}]$, and $\re (d^w_\tau) \propto \im [C_{eZ}^{33}]$, and the experimental measurements discussed in Sec.~\ref{sec:expStatus} can be readily used to probe the associated SMEFT dipole operators.

However, as we will see, even in this simplified setting, the SMEFT implies additional non-trivial correlations not only at tree-level, but also at higher orders. One striking effect is the SMEFT leading-logarithmic RG mixing of the $\tau$ dipole operators into the electron dipole operators, through intermediate mixings into bosonic operators~\cite{Jenkins:2013zja,Jenkins:2013wua,Alonso:2013hga}. For illustration, in the leading fixed-order approximation, this effect is given by 
\begin{equation}\begin{split}\label{eq:LLeX33toeX11}
    &C_{eW}^{11}(\mu)\approx \frac{y_e y_\tau}{(16\pi^2)^2}\Big[\Big(g_2^2 + \frac{9}{2}g_1^2\Big)C_{eW}^{33}(\Lambda) - \frac{3}{2}g_1 g_2 C_{eB}^{33}(\Lambda)\Big]\log^2\frac{\mu}{\Lambda} \,, \\[0.5em]
    &C_{eB}^{11}(\mu)\approx \frac{y_e y_\tau}{(16\pi^2)^2}\Big[3\Big(3g_1^2 + \frac{g_2^2}{2}\Big)C_{eB}^{33}(\Lambda) - \frac{9}{2}g_1 g_2 C_{eW}^{33}(\Lambda)\Big]\log^2\frac{\mu}{\Lambda} \,,
\end{split}\end{equation}
where $\Lambda$ is the NP scale and $\mu$ is a low energy scale, while $g_1,g_2$ are the SM gauge couplings and $y_e, y_\tau$ are the Higgs Yukawa couplings. As we will see in the subsequent sections, even with the relatively large suppression, these effects are phenomenologically relevant.

Similarly, the $\tau$ dipole operators mix into the modified Yukawa operator,
\begin{equation}
    \mathcal{O}_{eH}^{pr} = |H|^2 \bar{l}_p H e_r \,.
\end{equation}
As will be discussed in the next Section, this mixing can have important phenomenological effects due to contributions to the electron EDM through Barr-Zee diagrams. The leading effect is captured by the following approximate expression
\begin{equation}\label{eq:LLeX33toeH33}
    C_{eH}^{33}(\mu)\approx \frac{1}{16\pi^2}\big[3\big(3g_1^3 - g_1 g_2^2\big)C_{eB}^{33} + 9\big(g_1^2g_2 - g_2^3\big)C_{eW}^{33}\big]\log\frac{\mu}{\Lambda}\,.
\end{equation}
The above approximate expressions are written for illustrative purposes, whereas in the subsequent numerical analysis, we numerically solve the set of RG differential equations, resumming the leading logarithms to all orders.

\subsection{Complementary constraints}
The $\tau$ dipole operators in the SMEFT give rise to a set of complementary phenomenological effects that can be exploited to constrain the relevant Wilson coefficients. In addition to contributions to the $\tau$ weak dipole moments, these include direct contributions to the $\tau$ EM dipole moments, high-mass Drell-Yan production, the electron EDM, and precision measurements of partial $Z$ decay widths. In the following, we discuss these and other less relevant complementary constraints in more detail.

\subsubsection{Electric and magnetic dipole moments of leptons}\label{sec:dipoleMoments}
As shown in Eq.~\eqref{eq:CegammaCeZ}, the same dipole operators that contribute to the weak dipole moments of a lepton via $C_{eZ}$ also contribute to its EM electric and magnetic dipole moments via $C_{e\gamma}$~\cite{Aebischer:2021uvt}, but in perpendicular linear combinations. We parameterize the EM electric $d_\ell$ and magnetic $a_\ell$ dipole moments of a lepton $\ell$ via\footnote{Note that the normalization of the EM EDM differs from that of the wEDM defined in Eqs.~\eqref{eq:tauMultipoleME} and~\eqref{eq:multipoles}. Additionally, we emphasize that $a_\ell$ is the QED analogue to the weak magnetic dipole $\mu_\ell^w$, not the axial-vector form factor $a_\ell^w$.}
\begin{equation}
    \Gamma^{\mu(\gamma)}_\ell \supset \frac{e a_\ell}{2m_\ell}\sigma^{\mu\nu}q_\nu + i d_\ell\sigma^{\mu\nu}\gamma_5q_\nu \,,
\end{equation}
where
\begin{equation}
\begin{split}
    \mel{\ell^+(p_1)\ell^-(p_2)}{j_{\text{EM}}^\mu}{0} = \bar u(p_2)\Gamma_\ell^{\mu(\gamma)} v(p_1)\,,
    \end{split}
\end{equation}
with $j_{\text{EM}}^\mu = e \bar \ell \gamma^\mu \ell$ the EM current, and it is assumed that the matrix element is evaluated at $q^2=0$.
In this convention, the $C_{e\gamma}$ contributions are
\begin{equation}
\label{eq:ad_in_SMEFT}
\begin{split}
    &\rre \big(a_\ell\big) = \frac{4 m_\ell}{e}\rre\big[C_{e\gamma}^{\ell\ell}\big] \,,\\[0.5em]
    &\rre \big(d_\ell\big) = - 2\iim\big[C_{e\gamma}^{\ell\ell}\big]\,.
\end{split}
\end{equation}
Assuming only third-generation dipoles are generated at the NP scale, the most direct complementary constraints to the weak dipole moments of the $\tau$ are its electric and magnetic dipole moments. Currently, the best limits on the $C\!P$-violating electric dipole moment, $d_\tau$, come from 833 fb$^{-1}$ of $e^+e^-$ collisions at Belle~\cite{Belle:2021ybo}, giving
\begin{equation}
\label{eq:dtauexp}
    \rre\big(d_\tau\big) = (-0.62\pm0.63)\times 10^{-17} e\cdot\text{cm} \,.
\end{equation}
The most precise determination of the EM magnetic dipole moment comes from CMS~\cite{CMS:2024qjo} using $140\text{ fb}^{-1}$ of proton-proton collisions via $\gamma\gamma\to\tau^+\tau^-$, reading\footnote{
There is a determination of $a_\tau$ available from ATLAS using high-mass $\tau$ lepton pair production cross section measurements with $q^2\sim 10^2-10^3$ GeV~\cite{ATLAS:2025oiy}. This analysis treats $a_\tau$ as a short-distance WC, despite the highly off-shell photon involved in the process. This is in contrast to the CMS measurement, where protons radiate the initial-state photons in $\gamma\gamma\to\tau\tau$ which are nearly on-shell~\cite{CMS:2024qjo}. For this reason, we use the CMS determination in Eq.~\eqref{eq:CMSatau} for our analysis, despite the slightly improved value quoted in Ref.~\cite{ATLAS:2025oiy}.}
\begin{equation}
\label{eq:CMSatau}
    a_\tau = 0.0009^{+0.0032}_{-0.0031}\,.
\end{equation}
The same analysis also probed the electric dipole moment $d_\tau$, albeit with a worse precision than the Belle determination given above. ATLAS and DELPHI have also set bounds on $a_\tau$ from $\gamma\gamma\to\tau^+\tau^-$ in Pb-Pb collisions and $e^+e^-\to e^+e^-\tau^+\tau^-$, respectively, but with worse precision~\cite{ATLAS:2022ryk,DELPHI:2003nah}.
The standard model prediction of the $\tau$ EM magnetic dipole moment is~\cite{Eidelman:2007sb,Eidelman:2016aih,Keshavarzi:2019abf,DiLuzio:2024sps,Hoferichter:2025fea}
\begin{equation}
    a_{\tau,\text{SM}} = 1.177171(40)\times10^{-3} \,,
\end{equation}
where the leading errors come from hadronic uncertainties. Similar to $d_\tau^w$, $d_\tau$ is heavily suppressed in the SM and can be treated as zero when considering current and future experimental sensitivities.

The $\tau$ dipole operators mix into the electron dipole operators, as well as the $\mathcal{O}_{eH}^{33}$ operator due to one-loop RG evolution, as demonstrated in Eqs.~\eqref{eq:LLeX33toeX11} and \eqref{eq:LLeX33toeH33}. These in turn contribute to dipole moments of the electron ($e$EDM, $e$MDM), either directly at tree level, or through two-loop Barr-Zee diagrams (c.f. Eqs.~(25) and~(28) of Ref.~\cite{Brod:2022bww}). We find that, despite an additional loop suppression, the large anomalous dimension matrix entries describing the mixing of $C_{eX}^{33}\to C_{eH}^{33}$ for $X=W,B$ as well as the large logarithm $\sim\log^2(m_\tau^2/M_h^2)$ enhance the latter contribution, potentially leading to non-negligible bounds when compared to the $C_{eX}^{33}\to C_{eX}^{11}$ mixing. For this reason, we account for both contributions to electron dipole moments in our analysis. 

The $e$EDM is tightly constrained by precision measurements in polar molecules. The current most stringent bound comes from experiments using HfF$^{+}$ molecular ions, which report~\cite{Roussy:2022cmp}
\begin{equation}\label{eq:eEDMUpperLim}
    |d_e|<4.1\times 10^{-30}\, e \,\mathrm{cm}\,,
    \qquad90\%\,\mathrm{CL}\,.
\end{equation}
In this interpretation, electron-nucleon couplings have been neglected~\cite{Chupp:2017rkp,Ardu:2025rqy}, including contributions of $d_\tau$ to electron-nucleon CP-odd operators, which were shown to be subleading in Ref.~\cite{Ema:2021jds}.
On the other hand, interpretation of the anomalous magnetic moment of the electron $a_e$~\cite{VanDyck:1987ay,Hanneke_2008,Fan:2022eto,Aoyama:2019ryr} relies on the atomic interferometry determination of $\alpha$, which features a tension between experiments using different atoms~\cite{Parker_2018, Morel:2020dww} (see discussions in Refs.~\cite{Altmannshofer:2015qra, Kosnik:2025srw}). Nevertheless, we have checked that $a_e$ has a weak constraining power on $C_{eX}^{33}$ and has negligible impact on the subsequent numerical analyses.

\subsubsection{High-mass Drell-Yan tails}
The high-$p_T$ tails of the charged (CC) and neutral current (NC) Drell-Yan processes ${pp\to\ell \nu}$ and ${pp \to \ell\ell}$ are known to be excellent probes of flavorful SMEFT operators~\cite{Faroughy:2016osc, Greljo:2017vvb, Fuentes-Martin:2020lea, Allwicher:2022gkm, Greljo:2022jac,Grunwald:2023nli,Greljo:2023bab}. This holds especially for the semi-leptonic contact interactions, which have a favorable scaling with energy of $E^2/\Lambda^2$ at the amplitude level. The dipole operators have a slightly less favorable energy scaling of $vE/\Lambda^2$, however the constraints on them from these processes are still phenomenologically relevant. In this work we are primarily interested in the $\tau$ dipole operators $\mathcal{O}_{eW}^{33}$ and $\mathcal{O}_{eB}^{33}$, and we consider their tree-level contributions to the $pp\to \tau\tau$ and $pp\to \tau \nu$ processes. We use \texttt{HighPT}~\cite{Allwicher:2022mcg, Allwicher:2022gkm} to extract the necessary ingredients for our subsequent phenomenological analysis. This includes the experimental data from ATLAS on $pp\to \tau\tau$ and $pp \to \tau \nu$ differential cross sections at 139 fb$^{-1}$~\cite{ATLAS:2020zms, ATLAS:2024tzc}.

\subsubsection{$Z$ decays}

The coefficient $C_{eZ}$, defined in Eq.~\eqref{eq:CegammaCeZ}, is also accessible from the partial $Z\to\tau^+\tau^-$ decay width, given by
\begin{equation}\begin{split}
    \Gamma(Z\to\tau^+\tau^-) &= \Gamma_{\text{SM}}(Z\to\tau^+\tau^-) + \frac{M_Z}{4\pi}\Big[e m_\tau(g_R + g_L)\rre\big[C_{eZ}^{33}\big] + \frac{M_Z^2}{3}|C_{eZ}^{33}|^2\Big]\,,
\end{split}\end{equation}
where in the NP contribution, we neglect terms $\mc{O}(m_\tau^2/M_Z^2)$ and $g_{L(R)}$ are given in Eq.~\eqref{eq:gRgL}. The partial width itself is strongly dependent on the choice of the electroweak parameter input scheme which receives tree-level corrections from SMEFT WCs~\cite{Biekotter:2025nln}. However, the ratio
\begin{equation}
    R_\ell = \frac{\Gamma(Z\to\text{hadrons})}{\Gamma(Z\to\ell^+\ell^-)}\,,
\end{equation}
is largely independent of this scheme choice. In order to avoid non-negligible QCD corrections, we opt to use the double ratio
\begin{equation}\label{eq:doubleRatio}
    R_{\tau/\mu} \equiv \frac{R_\mu}{R_\tau} = 1 + \frac{3}{2e^2(g_R^2 + g_L^2)}\Big[4em_\tau(g_R + g_L)\rre \big[C_{eZ}^{33}\big] + \frac{4M_Z^2}{3}|C_{eZ}^{33}|^2\Big]\,,
\end{equation}
where the theory uncertainty on the standard model part of the double ratio is negligible~\cite{Dubovyk:2019szj}. Note that neither term in the brackets of Eq.~\eqref{eq:doubleRatio} can necessarily be neglected, despite the first term being suppressed by a factor of $e m_\tau$ and the second being suppressed by an additional power of the SMEFT WC. In fact, if $\rre [C_{eZ}^{33}]\sim 1/\Lambda$, one sees that the terms are competitive when
\begin{equation}
    e(g_R + g_L)\frac{m_\tau}{M_Z}\sim\frac{1}{3}\frac{M_Z}{\Lambda}\,.
\end{equation}
For $\Lambda\ll 50$ TeV, the term quadratic in $C_{eZ}$ dominates and the linear term can be neglected. However, when $\Lambda\sim 50$ TeV, the two terms are of comparable size, and above this scale, the linear term becomes dominant.

The PDG values for the width ratios are~\cite{ALEPH:2005ab}
\begin{equation}
    R_\mu = 20.785\pm0.033\,,\qquad R_\tau = 20.764\pm0.045 \,,
\end{equation}
giving
\begin{equation}\label{eq:RtauRmuExp}
    R_{\tau/\mu} = 1.0010\pm 0.0027\,.
\end{equation}

\subsubsection{Other constraints}

Additional constraints can be placed on the $\tau$ dipole operators $C_{eX}^{33}$ from various observables, albeit with less sensitivity compared to the constraints discussed above. For example, the branching ratio of $W\to\tau\nu$ measured at LEP~\cite{ALEPH:2013dgf}, $\mrm{Br}(W \to \tau\nu) = (11.38\pm 0.21)\%$ and recently also by CMS, $\mrm{Br}(W \to \tau\nu) = (10.77\pm 0.05\pm 0.21)\%$~\cite{CMS:2022mhs} can be sensitive to $C_{eW}^{33}$. Similarly, the $C\!P$-even and $C\!P$-odd observables in $h\to \tau\tau$ (see e.g.~\cite{ATLAS:2016neq, ATLAS:2022akr, CMS:2021sdq}) can be sensitive through the mixing of $C_{eX}^{33}$ into $C_{eH}^{33}$ discussed in Sec.~\ref{sec:eftSetup}, as well as direct one-loop contributions at the EW scale. We have explicitly checked that the observables mentioned here lead to significantly weaker constraints compared to the observables described above, and do not discuss them further.

\subsection{Phenomenology}\label{sec:pheno}
\subsubsection{Numerical setup}
For our phenomenological analysis we use the open source Python package \texttt{jelli} (JAX-based EFT likelihoods)~\cite{Smolkovic:2026cba, jelli}, which implements a general framework for fast and differentiable EFT likelihoods. To this end, we implement the theory predictions of all considered observables in the Polynomial Observable Prediction exchange format (\texttt{POPxf})~\cite{Brivio:2025mww}, in the SMEFT \texttt{Warsaw} basis as defined in \texttt{WCxf}~\cite{Aebischer:2017ugx}. This allows us to automatically include the leading-logarithmic RG evolution of the SMEFT operators due to the interface between \texttt{jelli} and \texttt{rgevolve}~\cite{Smolkovic:2026cba, rgevolve}. The latter package implements evolution, translation, and matching matrices, computed using \texttt{wilson}~\cite{Aebischer:2018bkb}, for different bases of various EFTs, following the conventions in \texttt{WCxf}. As the provided evolution matrices have been obtained by fully solving the RG differential equations, the full resummation of leading logarithms is captured. We have explicitly cross-checked that the results obtained in this way agree well with the leading fixed-order expressions provided in Sec.~\ref{sec:eftSetup} in the regime where the logarithms remain $\lesssim\mathcal{O}(1)$. Finally, the experimental data has also been implemented in the format appropriate for use in \texttt{jelli}.

\subsubsection{Results}
\label{sec:mainResults}
\begin{figure}[t]
     \centering
     \begin{subfigure}[b]{0.45\textwidth}
         \centering
         \includegraphics[width=\textwidth]{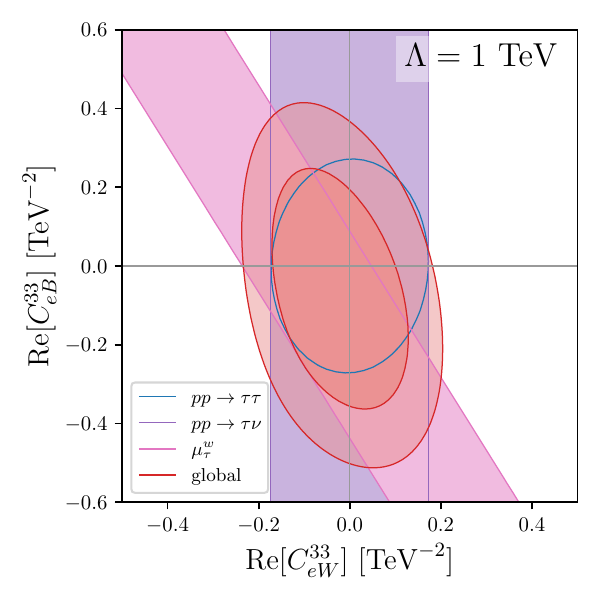}
     \end{subfigure}~
          \begin{subfigure}[b]{0.45\textwidth}
         \centering
         \includegraphics[width=\textwidth]{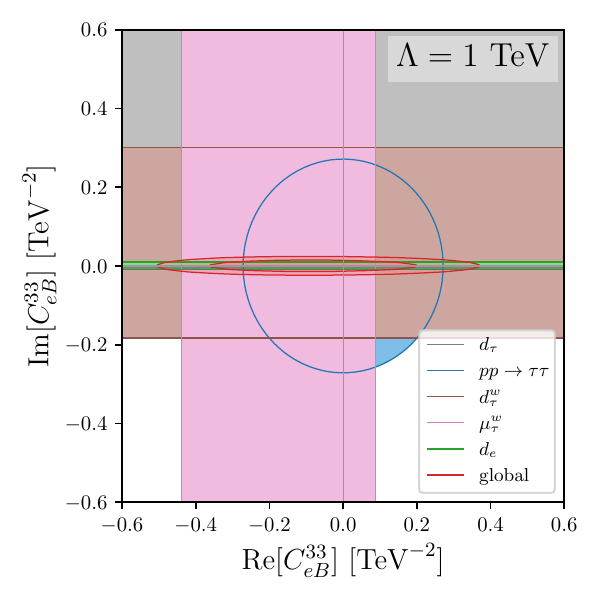}
     \end{subfigure}
     \\
     \begin{subfigure}[b]{0.45\textwidth}
         \centering
         \includegraphics[width=\textwidth]{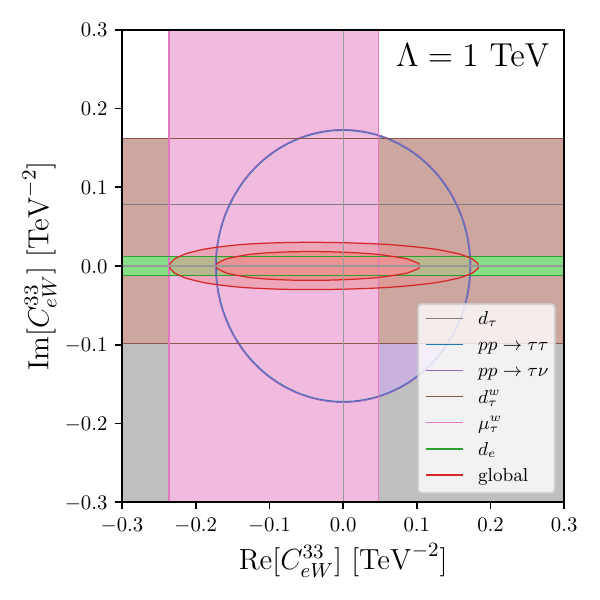}
     \end{subfigure}~
     \begin{subfigure}[b]{0.45\textwidth}
         \centering
         \includegraphics[width=\textwidth]{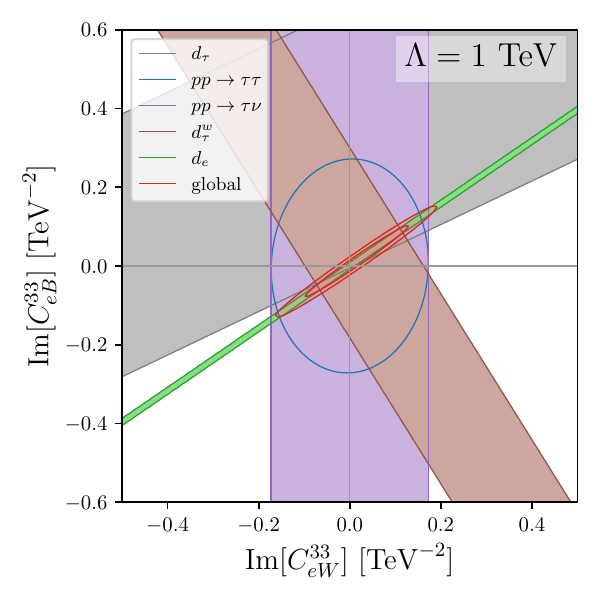}
     \end{subfigure}
        \caption{The $68\%$ CL (and $95\%$ CL in the global fit) constraints on four different scenarios involving the SMEFT $\tau$ dipole operators considered in this work, with the NP scale set to $\Lambda=1~\mathrm{TeV}$. See surrounding text for a more detailed discussion.}
        \label{fig:EFTresults}
\end{figure}

The main results of our analysis of current bounds are presented in Fig.~\ref{fig:EFTresults}\footnote{We use the plotting functions implemented in \texttt{flavio}~\cite{Straub:2018kue} to generate the figures.}. For all the plots, we set the UV matching scale to $\Lambda=1~\mathrm{TeV}$ and run the Wilson coefficients down to the electroweak scale where most of our observable predictions are computed, with the exception of Drell-Yan, for which a renormalization scale of $1~\mathrm{TeV}$ is used. The plots show the 68\% CL contours of $\Delta \chi^2(C) = \chi^2(C)-\chi^2_\mathrm{min}$, computed with the appropriate degrees of freedom, and where $\chi^2(C)\equiv -2 \log L(C)$, with likelihood function, $L$. For the global fits, the 95\% CL contours are also shown. Each plot corresponds to a particular $2$D scenario, where the plotted WCs are evaluated at $\Lambda$ and all remaining WCs are set to zero. Any observable not shown on a particular plot turns out to have comparatively weak or no sensitivity to that particular scenario. In the following, we discuss each plot in the figure in more detail.

In Fig.~\ref{fig:EFTresults} (upper left) we show the constraints in the $(\re [C_{eW}^{33}], \re [C_{eB}^{33}])$ plane, corresponding to a purely real assumption about the $\tau$ dipole operators in the SMEFT. Both NC (blue) and CC (purple) Drell-Yan processes provide important constraints in this plane, the CC process being insensitive to the $\re [C_{eB}^{33}]$ direction, while the NC process alone cannot independently constrain both coefficients. The weak magnetic dipole moment of the $\tau$, as shown in pink, provides precise constraints in this plane, albeit also suffering from a flat direction (related to the rotation by the weak mixing angle). Although Drell-Yan data alone could close the fit in this plane, $\mu_\tau^w$ brings important information, as can be seen from the close alignment of the global fit in red. Finally, we comment that the constraint from the anomalous magnetic moment of the $\tau$ would in principle lead to a near\footnote{The exact perpendicularity is broken by RG effects.} perpendicular constraint to that from the weak magnetic moment, however at current experimental precision the constraint is not relevant and does not appear in the plot, nor does it significantly impact the global fit.

Additionally, in Fig.~\ref{fig:EFTresults} we show the constraints in the $(\re [C_{eB}^{33}], \im [C_{eB}^{33}])$ (upper right)  and $(\re [C_{eW}^{33}], \im [C_{eW}^{33}])$ (lower left) planes. These scenarios are less natural as they assume the presence of just one of these operators in the UV (see Sec.~\ref{sec:uvmodels}), with complex Wilson coefficients. Nevertheless, they are instructive. In both cases, the leading constraint in the respective imaginary direction comes from the electron EDM (green), through the RG operator mixing effects discussed in Sec.~\ref{sec:eftSetup}. The weak electric dipole moment of the $\tau$ (brown) can also constrain this direction, however with less sensitivity. In both planes the Drell-Yan constraints (blue and purple) are relatively important, however also the weak magnetic dipole moment of the $\tau$ (pink) can efficiently constrain the perpendicular direction. Finally, we show constraints from the $\tau$ EDM in gray, which are slightly asymmetric, as expected from Eq.~\eqref{eq:dtauexp}, but not particularly relevant for the global fit at the current level of experimental precision.

In Fig.~\ref{fig:EFTresults} (lower right), we show the constraints in the $(\im [C_{eW}^{33}], \im [C_{eB}^{33}])$ plane, corresponding to a purely imaginary UV scenario. This plot most importantly illustrates the complementarity between electron EDM (green) and $\tau$ wEDM (brown) constraints: accounting only for RG mixing in SMEFT, these lead to almost perpendicular constraints. Even though the electron EDM is more sensitive in absolute terms, it alone can not lift the degeneracy between the two operators. As can be deduced from the global fit in red, the wEDM of the $\tau$ alone is capable of breaking the degeneracy and closing the fit, with NC and CC Drell-Yan  (blue and purple) playing a less important role. Finally, we also note the important perpendicularity between the $\tau$ EDM (gray) and wEDM (brown).

We reiterate that, although the contribution to $d_e$ from the mixing of $C_{eX}^{33}$ into $C_{eH}^{33}$ and the subsequent matching onto the relevant LEFT operator are implemented exactly, we only account for the leading-logarithmic running of $C_{eX}^{33}$ into $C_{eX}^{11}$ in the SMEFT. In our fits, we neglect the contribution from QED running of $C_{e\gamma}^{11}$ below the EW scale, as well as the direct matching contribution from two-loop Barr-Zee-like diagrams and one-loop matching onto four-lepton operators which subsequently mix into $C_{e\gamma}^{11}$ at one-loop below the EW scale. The former effect is anticipated to be very small, but the latter two effects can in principle contribute sizable corrections, particularly the last, which is expected to be enhanced by large logarithms $\sim\log m_\tau^2/M_Z^2$. To our knowledge, these effects have not been explicitly calculated, and although we do not expect them to dramatically change the order of magnitude of the bounds placed in the $(\im [C_{eB}^{33}],\im [C_{eW}^{33}])$ plane, they could change the rotation of the resulting flat direction. This could, in principle, affect our conclusions of the complementarity between $d_\tau^w$ and $d_e$ in the unlikely fine-tuned scenario where these additional effects would rotate the flat direction of the $e$EDM to be aligned with $C_{eZ}^{33}$, motivating a more dedicated computation which is beyond the scope of this current work.

\begin{table}[t]
  \centering
  \caption{Results of the global $4$-dimensional fit of real and imaginary parts of the SMEFT $\tau$ dipole operators, in the Gaussian approximation. Central values and standard deviations are given in the left part, while the correlation matrix is provided in the right part.}
  \label{tab:gaussian_fit}
  \begin{tabular}{lr|lrrrr}
    \toprule
    & & & $\mathrm{Re}[C_{eB}^{33}]$ & $\mathrm{Re}[C_{eW}^{33}]$ & $\mathrm{Im}[C_{eB}^{33}]$ & $\mathrm{Im}[C_{eW}^{33}]$ \\
    \midrule
    $\mathrm{Re}[C_{eB}^{33}]$ & $-0.068 \pm 0.217$ & $\mathrm{Re}[C_{eB}^{33}]$  & 1 & -0.462 & -0.008 & -0.009 \\
    $\mathrm{Re}[C_{eW}^{33}]$ & $-0.029 \pm 0.111$ & $\mathrm{Re}[C_{eW}^{33}]$  &  & 1 & 0.000 & 0.000 \\
    $\mathrm{Im}[C_{eB}^{33}]$ & $0.008 \pm 0.062$ & $\mathrm{Im}[C_{eB}^{33}]$  &  &  & 1 & 0.988 \\
    $\mathrm{Im}[C_{eW}^{33}]$ & $0.010 \pm 0.078$ & $\mathrm{Im}[C_{eW}^{33}]$  &  &  &  & 1 \\
    \bottomrule
  \end{tabular}
\end{table}

Finally, we perform a combined $4$-dimensional global fit of the real and imaginary parts of $C_{eB}^{33}$ and $C_{eW}^{33}$, again at the scale of $\Lambda=1~\mathrm{TeV}$. We report the results in a compact form in Tab.~\ref{tab:gaussian_fit}. These consist of the best fit values, obtained by maximizing the global likelihood. At that point in the parameter space, the exact Hessian matrix is evaluated, from which the Gaussian covariance matrix is computed and the associated standard deviations and the correlation matrix are extracted. As already anticipated from the above examples, the imaginary parts are better constrained than the real parts. This is primarily due to the electron EDM, which is also the reason behind the strong correlation between $\im [C_{eB}^{33}]$ and $\im[C_{eW}^{33}]$ that could already be observed in Fig.~\ref{fig:EFTresults} (lower right). Nevertheless, the exact degeneracy is broken by complementary constraints, among which the most important one is the $\tau$ wEDM. The real parts $\re [C_{eB}^{33}]$ and $\re[C_{eW}^{33}]$ are somewhat anti-correlated, as could already be observed on Fig.~\ref{fig:EFTresults} (upper left), as the strongest constraints come from the $\tau$ wMDM. Here, the complementary constraints from Drell-Yan data play a crucial role to break the flat direction.

\subsubsection{Implications for UV models}\label{sec:uvmodels}
The results of the previous section can be used to constrain UV completions of the SM which generate $C_{eX}^{33}$ when integrating out heavy degrees of freedom and matching onto the SMEFT. However, it is important to discuss the general validity and assumptions of these constraints when considering specific models.

First, it is very challenging to generate just one of $C_{eB}$ or $C_{eW}$, but not the other, from a concrete UV model, since the structure of the dipole operators in SMEFT naturally couples left- and right-handed SM degrees of freedom. This implies that either left- and right-handed SM fields themselves appear inside loops in the matching procedure, or any high-energy degrees of freedom which generate these operators will be charged under both $SU(2)$ and $U(1)$ gauge groups in order to couple to the SM fields. Thus the two off-diagonal plots in Fig.~\ref{fig:EFTresults} should be taken as illustrative only, since they assume that only one of the two WCs are non-zero.

Additionally, and perhaps more critically, many more WCs other than $C_{eX}^{33}$ will be generated from a generic UV model. This is a particularly important point, since dimension-six dipoles in the SMEFT are generated at the loop-level, so one would not typically expect new physics signals to first appear from $\mathcal{O}_{eX}$, but instead from operators which are generated at tree-level. That said, many models -- particularly those which couple mainly to the third generation -- produce SMEFT operators at tree-level which are difficult to constrain in experiments, making loop-generated operators competitive~\cite{Allwicher:2023shc,Maura:2024zxz, Arroyo-Urena:2017sfb}. As a particularly simple example, consider a third-generation-coupled leptoquark $\Pi_7\sim(3, 2)_{7/6}$ (using the notation of Ref.~\cite{deBlas:2017xtg}, commonly known as $R_2$) with the following interaction Lagrangian
\begin{equation}\label{eq:R2Lagrangian}
    \mathcal{L} \supset -\big(y_{\Pi_7}^{lu}\big)_{33}\Pi^\dag_7 i\sigma_2\bar l_3^T\,u_3 - \big(y_{\Pi_7}^{qe}\big)_{33}\Pi^\dag_7 \bar e_3 q_3 + \text{h.c.} \,.
\end{equation}
Integrating out the heavy leptoquark generates the SMEFT operators
\begin{equation}\begin{split}
    &\mathcal{O}_{lu}^{3333} = \big(\bar l_3\gamma^\mu l_3\big)\big(\bar u_3\gamma_\mu u_3\big)\,, \\[0.5em]
    &\mathcal{O}_{lequ}^{(1),3333} = \big(\bar l^j_3 e_3\big)\epsilon_{jk}\big(\bar q_3^k u_3\big)\,,
\end{split}
\qquad
\begin{split}
    &\mathcal{O}_{qe}^{3333} = \big(\bar q_3\gamma^\mu q_3\big)\big(\bar e_3\gamma_\mu e_3\big)\,, \\[0.5em]
    &\mathcal{O}_{lequ}^{(3),3333} = \big(\bar l^j_3 \sigma_{\mu\nu} e_3\big)\epsilon_{jk}\big(\bar q_3^k \sigma^{\mu\nu} u_3\big)\,,
\end{split}\end{equation}
at tree-level. All four of these operators are very difficult to bound using tree-level observables:
$\mathcal{O}_{l u}^{3333}$ can only appear in top/$\tau$ or top/$\nu_\tau$ observables, $\mathcal{O}_{q e}^{3333}$ enters $B\to K^{(*)}\tau^+\tau^-$ and $b\bar b\to \tau^+\tau^-$ which are poorly measured and heavily PDF-suppressed, respectively, and both $l e qu$ operators will contribute to $t\to b\tau\nu$ decays, but only interfere with the SM with a chiral suppression $\sim m_\tau$. In this case then, loop-suppressed operators/observables, such as those which contribute in the $Z\to\tau\tau$ observables considered in this work, can result in competitive, if not better, constraints.

From the results of Section~\ref{sec:mainResults}, it is clear that the weak dipole moments of the $\tau$ give strong constraining power when considered on equal footing with other observables such as high-$p_T$ Drell-Yan, electroweak boson decays, and EM $\tau$ dipoles. In the very simple model in Eq.~\eqref{eq:R2Lagrangian}, the dipoles which are generated would appear as lines in the top-left and bottom-right plots of Fig.~\ref{fig:EFTresults}~\cite{Dorsner:2016wpm}, and would be mostly constrained by $\mu_\tau^w$ for the former, and the $e$EDM for the latter, unless these lines fall along the flat directions corresponding to either of these observables. In this case, complementary observables such as Drell-Yan can be used to close the fit in the real-real plane, while $d_\tau^w$ is the strongest complementary constraint to the $e$EDM in the imaginary-imaginary plane. Of course, slightly more complicated models with additional couplings and/or masses (e.g.~from additional leptoquarks or vector-like fermions) can break the correlations between $C_{eW}^{33}$ and $C_{eB}^{33}$, in which case, the weak dipoles serve as vital observables for closing the fit and placing the best possible combined constraints in both planes. That said, placing a complete set of constraints on such models would require a comprehensive analysis of one-loop observables generated in the SMEFT. However, this qualitative discussion suggests that measurements of EM and weak dipole moments can contribute crucial constraints for models with suppressed or experimentally challenging tree-level processes.

As a final comment, when considering TeV-scale NP models with an EFT analysis, one must take into account the validity of the EFT itself, particularly in observables such as high-$p_T$ Drell-Yan tails~\cite{Allwicher:2024mzw,Greljo:2022jac}. In fact, for models which generate operators relevant for Drell-Yan observables at the loop-level, perturbative unitarity can place stronger bounds than high-$p_T$ tails, especially in high-energy bins~\cite{Cohen:2021gdw}, though for dipole operators themselves, it has been shown that lower-energy bins ($\lesssim 1$ TeV) dominate the constraints~\cite{Allwicher:2022gkm}. On the other hand, weak dipole moments are truly low-energy observables and do not suffer the same potential pitfalls in an EFT analysis, as long as the new physics scale is far enough separated from the electroweak scale for a valid operator product expansion.

\section{Future prospects}
\label{sec:future}

In this Section, we first consider the future prospects of measuring the SM-predicted value of the $\tau$ weak magnetic moment, discussed in Sec.~\ref{sec:smRes}, at FCC-$ee$, paying particular attention to the fact that such a measurement would be systematics dominated. Secondly, we consider the projected sensitivities to heavy NP effects of the leading observables discussed in Sec.~\ref{sec:NP}, considering both HL-LHC~\cite{Apollinari:2015wtw,ZurbanoFernandez:2020cco} and FCC-$ee$ experimental programs. Special emphasis is given to assumptions made about systematic uncertainties.

\subsection{Prospects for measuring the Standard Model}\label{sec:measureSMDipole}
\label{sec:future_prospects_sm}
We first consider the future prospects of measuring the SM values of the $\tau$ weak magnetic dipole moment, discussed in Sec.~\ref{sec:smRes}, focusing on its real part (the prospects for the imaginary part are similar). To illustrate the potential constraining power of FCC-$ee$ we use the following procedure. First, we split the experimental covariance matrix into the statistical and systematic part as
\begin{equation}
    \Sigma = \Sigma_\mathrm{stat} + \Sigma_\mathrm{syst}\,,
\end{equation}
where $\Sigma_\mathrm{stat}$ and $\Sigma_\mathrm{syst}$ are reported by ALEPH~\cite{ALEPH:2002kbp}. For the statistical term, we assume scaling of uncertainties with a luminosity improvement factor. FCC-$ee$ is reported to be able to deliver about $r_\mathrm{stat} = 2\times 10^5$ times the integrated luminosity at the $Z$ pole compared to the one produced by LEP~\cite{FCC:2018byv}. As statistical uncertainties scale as $\sigma \propto 1/\sqrt{N}$ and $N\propto L$, we rescale $\Sigma_\mathrm{stat}$ with a factor $1/r_\mathrm{stat}$.

\begin{figure}[t]
     \centering
     \begin{subfigure}[b]{0.7\textwidth}
         \centering
         \includegraphics[width=\textwidth]{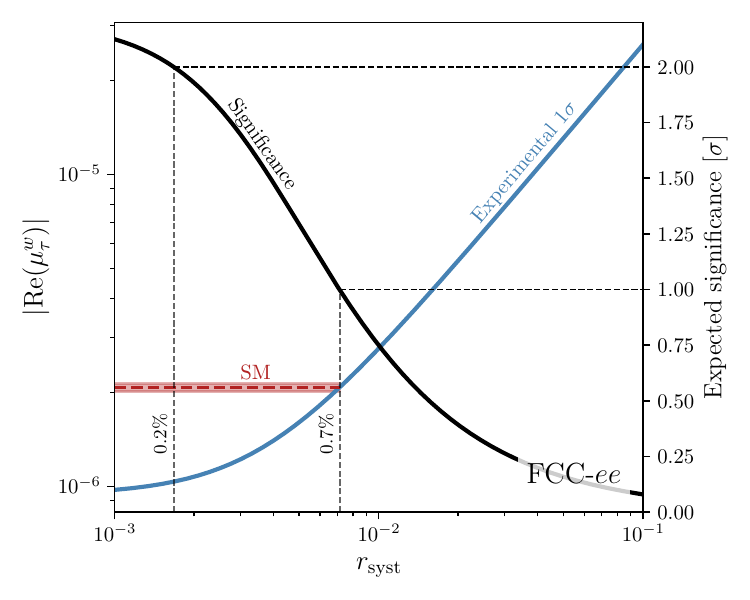}
   \end{subfigure}
        \caption{
        The prospects of measuring the SM-predicted value of the real part of the $\tau$ weak dipole moment, $\re(\mu_\tau^w)$, at FCC-$ee$. The projected experimental uncertainty is shown in blue (left axis), while the total projected significance of SM observation is shown in black (right axis), both as functions of the relative improvement factor in systematic uncertainties, $r_\mathrm{syst}$, defined in Eq.~\eqref{eq:SigmaFCC}. See surrounding text for a detailed discussion.
        }
        \label{fig:fcc-wmdm-projection}
\end{figure}

The expected improvement in statistics is immense and we anticipate the measurements considered here to be heavily systematics dominated at FCC-$ee$, see e.g.~\cite{FCC:2018byv}. A dedicated FCC-$ee$ analysis with full MC including detector simulation and accounting for reconstruction efficiencies would have to be performed in order to quantify these effects properly, which is beyond the scope of this work. Nevertheless, we take a pragmatic approach and assume a relative improvement of the systematic uncertainties, then consider and comment on various values of the relative systematic uncertainty rescaling factor $r_\mathrm{syst}$. The combined projected covariance matrix is therefore
\begin{equation}
\label{eq:SigmaFCC}
    \Sigma^\mathrm{FCC} = \frac{1}{r_\mathrm{stat}} \Sigma_\mathrm{stat} + {r_\mathrm{syst}^2} \Sigma_\mathrm{syst}\,.
\end{equation}

In Fig.~\ref{fig:fcc-wmdm-projection} (left vertical axis) we show the dependence of the values of $|\re(\mu_\tau^w)|$ potentially probed at FCC-$ee$ as a function of the systematics relative improvement factor $r_\mathrm{syst}$. The projected total $1\sigma$ experimental uncertainty is shown with the blue curve, and the SM prediction including the $1\sigma$ theory uncertainty band is shown in red. The blue curve can also be straightforwardly reinterpreted as the projected upper limit on $|\re(\mu_\tau^w)|$, assuming no signal is observed. On the same plot, with the right vertical axis, we show the dependence of the expected significance of the SM observation in units of $\sigma$ (black curve), where a measurement of the SM-predicted value is assumed, and both the theoretical and experimental uncertainties are accounted for, although the theory uncertainties are negligible. The right-most value of $r_\mathrm{syst}$ implies a somewhat conservative reduction in systematic uncertainties by a factor $10$ with respect to ALEPH, while the left-most value corresponds to a much stronger improvement by a factor $1000$, which is possibly unrealistic. Indeed, Tab.~3.1 of Ref.~\cite{FCC:2018byv} -- in particular the entry for the $\mathrm{A}_\mathrm{FB}^{\mathrm{pol},\tau}$ observable, for which the experimental error is dominated by $\tau$ polarization and decay physics -- suggests that an improvement of at least $r_\mathrm{syst}=4\%$ is expected, assuming the current total uncertainties are systematics dominated. We nevertheless show this wide plot range for two reasons: firstly, we can observe the flattening of both the projected experimental uncertainties and the expected significance for small values of $r_\mathrm{syst}$, at which stage the measurement starts becoming statistics-dominated, illustrating that an immense improvement in systematics is needed to reach the full statistical potential of FCC-$ee$. Secondly, we can deduce the improvement in systematics required to probe the SM at $1$ or $2\sigma$, which we show with dashed black vertical and horizontal lines as $r_\mathrm{syst}\approx0.7\%$ and $0.2\%$, respectively. Our results set a clear target for systematics improvement required to be able to start probing the SM values of the weak magnetic moment of the $\tau$, and only a dedicated experimental sensitivity study can tell if such an improvement is attainable.

\subsection{Prospects for probing heavy New Physics}
\label{sec:future_prospects_bsm}

Having discussed the prospects for probing the SM-predicted value of the weak magnetic moment of the $\tau$, we now turn to future prospects for the constraining power of the main observables on the $\tau$ dipole operators in the SMEFT. From the results discussed in Sec.~\ref{sec:mainResults} we can conclude that both the wEDM and the wMDM of the $\tau$ provide crucial constraints on the $\tau$ dipole operators in the SMEFT, as can be seen most clearly in the upper-left and lower-right plots of Fig.~\ref{fig:EFTresults}. Although Drell-Yan processes do provide important constraints, the weak moments of the $\tau$ are more sensitive, and are crucial especially in closing the fit in the imaginary plane of these operators, in complementarity to the electron EDM. Additionally, we discuss the projected sensitivity of $Z\to\tau\tau$ via $R_{\tau/\mu}$, defined in Eq.~\eqref{eq:doubleRatio}, at FCC-$ee$, and of the $\tau$ magnetic dipole moment, $a_\tau$. For all the projections we consider expected bounds, assuming the SM predictions of all observables as central experimental values, with reasonable assumptions on theoretical and experimental uncertainties that differ for each observable and are discussed in the following paragraphs.

Considering the weak dipole moments of the $\tau$ at FCC-$ee$, we follow the procedure described in Sec.~\ref{sec:future_prospects_sm}, accounting for the luminosity improvement factor in the statistical covariance matrix, and a relative systematics improvement factor in the systematic one. We emphasize again that we expect the measurement to be systematics dominated, hence an assumption on this improvement plays a key role in assessing the future sensitivity. As already mentioned in the previous subsection, an improvement of at least $r_\mathrm{syst}=4\%$ is reasonably expected from similar analyses by the FCC collaboration~\cite{FCC:2018byv}. Nevertheless, to be even more conservative, we only assume an improvement $r_\mathrm{syst}=10\%$ in this section. As we will see, even this (possibly pessimistic) assumption leads to an impressive improvement in the sensitivity to heavy NP.

For Drell-Yan processes we follow the procedure described in Ref.~\cite{Allwicher:2022mcg} and assume a luminosity-based rescaling of the expected number of events and the associated statistical uncertainty in order to obtain the projected bounds for HL-LHC, assuming the integrated luminosity of $3~\mathrm{ab}^{-1}$. In addition, we assume a similar improvement in systematics as statistics, which we consider to be an optimistic projection, and should act as a rough best estimate of the future sensitivity.

Similarly, we consider the potential future improvement in measuring the $\tau$ anomalous magnetic moment, $a_\tau$. Currently the best constraint comes from CMS~\cite{CMS:2024qjo}, as given in Eq.~\eqref{eq:CMSatau}. As a rough estimate of the future sensitivity at HL-LHC, we again perform a luminosity-based rescaling, assuming the same improvement in statistics and systematics, acting as an optimistic estimate. This results in a projected combined experimental uncertainty of $7\times 10^{-4}$. The same quantity can also be probed at the Tera-$Z$ run of FCC-$ee$, and as an example we take the $e^+e^-\to \tau\tau \gamma$ process at the $Z$-pole, which was considered in the past by L3~\cite{L3:1998gov} and OPAL~\cite{OPAL:1998dsa} experiments at LEP. Again we expect such measurement to be systematics dominated, and conservatively assuming $r_\mathrm{syst}=10\%$, the resulting projected experimental uncertainty is at the level of $2\times10^{-3}$, not reaching the one potentially achievable by CMS at HL-LHC. In principle, $a_\tau$ can also be measured at FCC-$ee$ through $e^+e^-\to e^+ e^-\tau^+\tau^-$ either at the $Z$-pole or $WW$ threshold. Currently, the best bounds from this process come from the DELPHI collaboration using $650 \text{ pb}^{-1}$ of LEP-II data at center of mass energies $183\text{ GeV} < \sqrt{s} < 208\text{ GeV}$, giving $a_\tau = -0.018\pm 0.017$~\cite{DELPHI:2003nah}. An improvement in systematics of at least a factor of $10$ would be required in order to be competitive with the projection of CMS at HL-LHC discussed above, motivating a dedicated experimental sensitivity analysis. Finally, it has been shown in Refs.~\cite{Bernabeu:2007rr,Bernabeu:2008ii,Gogniat:2025eom,Crivellin:2021spu} that linear combinations of transverse and longitudinal $\tau$ asymmetries in $e^+e^-\to \tau^+\tau^-$ from which $a_\tau$ can be extracted are relatively insensitive to systematic uncertainties, and can potentially be measured at Chiral Belle~\cite{Roney:2025oj,USBelleIIGroup:2022qro} to a precision of $10^{-5}$. However, since the status of Chiral Belle is currently unclear, we do not include this projection of $a_\tau$. That said, including this projected precision on $a_\tau$ would not change our general conclusions about the weak dipole moments, since they remain highly complementary.

Finally, FCC-$ee$ should also improve on the measurement of the $Z$ partial decay rates, and subsequently the $R_{\tau/\mu}$ ratio defined in Eq.~\eqref{eq:doubleRatio}. Following Ref.~\cite{FCC:2018byv}, the systematics on the individual $R_\ell$ ratios are expected to improve by a factor of about $20-100$. To be consistent with the projections of $\tau$ weak dipole moments, we take a conservative approach and compute the corresponding expected improvement in systematics of $R_{\tau/\mu}$ using ${r_\mathrm{syst}\approx5\%}$, following the definition in Eq.~\eqref{eq:SigmaFCC}.

\begin{figure}[t]
     \centering
     \begin{subfigure}[b]{0.45\textwidth}
         \centering
         \includegraphics[width=\textwidth]{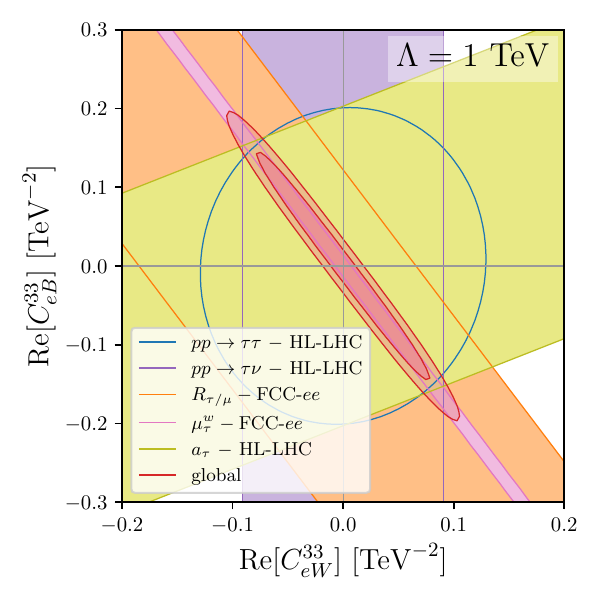}
     \end{subfigure}~
     \begin{subfigure}[b]{0.45\textwidth}
         \centering
         \includegraphics[width=\textwidth]{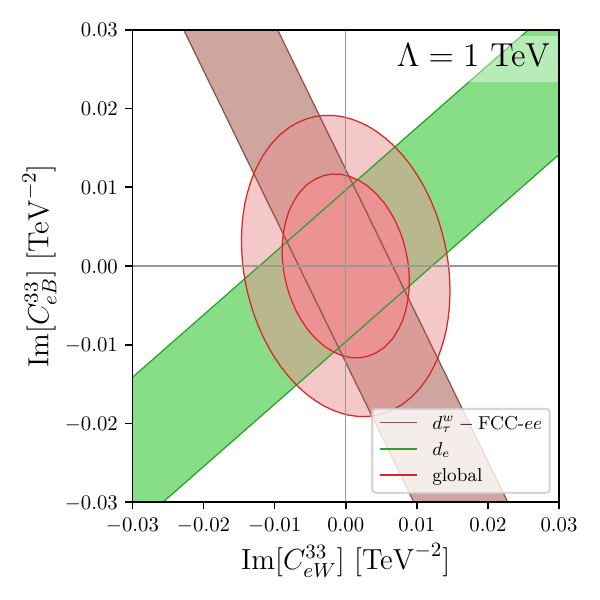}
     \end{subfigure}
        \caption{
        The projected $68\%$ CL (and $95\%$ CL in the global fit) constraints on two different scenarios involving the SMEFT $\tau$ dipole operators considered in this work, with the NP scale set to $\Lambda=1~\mathrm{TeV}$. The expected sensitivities of observables at either HL-LHC (optimistic assumption on systematics) or FCC-$ee$ (conservative assumption on systematics) are shown. See surrounding text for a more detailed discussion.}
        \label{fig:EFTresults-projected}
\end{figure}

To demonstrate the future reach of these observables, we revisit two scenarios already discussed in Sec.~\ref{sec:mainResults}, namely the purely real and purely imaginary planes of the $C_{eW}^{33}$ and $C_{eB}^{33}$ Wilson coefficients. The results are shown in Fig.~\ref{fig:EFTresults-projected} left and right, respectively. First focusing on the purely real plane, the measurement of the weak magnetic moment $\mu_\tau^w$ (pink) can reach an impressive sensitivity even with the pessimistic assumption on the improvement of systematics. This is the most sensitive constraint in this plane, and the $Z$ partial decay widths ($R_{\tau/\mu}$, orange) do not reach the same level of sensitivity. Complementary information is provided first by Drell-Yan processes at HL-LHC, but also by the anomalous magnetic moment of the $\tau$, $a_\tau$, as projected to HL-LHC (measured in $pp$ collisions), which leads to important perpendicular constraints in this plane. 

Finally, considering the purely imaginary plane in Fig.~\ref{fig:EFTresults-projected} (right), only two constraints remain important at the level of sensitivity shown in the plot: $d_\tau^w$ and $d_e$. Note that we did not project $d_e$ due to the fact that the theoretical prediction is incomplete, despite future experiments which can improve the bounds by up to an order of magnitude~\cite{Ng_2022,2019APS..DMPC02003N}. Similarly, we did not project the constraints due to $d_\tau$, as we do not expect them to be competitive. The constraint from the wEDM of the $\tau$, $d_\tau^w$, are shown in brown and reach an impressive sensitivity even with a conservative estimate of the systematics improvement. The constraints due to the electric dipole moment of the electron, $d_e$, are shown in green. The two constraints, under these assumptions, reach a similar level of sensitivity to the $\tau$ dipole operators in the SMEFT, and completely close the fit. 

\section{Conclusions}
\label{sec:conc}
In this work, we studied the weak magnetic and electric dipole moments of the $\tau$ lepton as potential precision probes of the SM and observables sensitive to heavy NP, achieving three complementary goals: an updated SM prediction for $\mu_\tau^w$ with a careful assessment of theoretical uncertainties, a comprehensive SMEFT analysis of the $\tau$ dipole operators incorporating all relevant complementary constraints, and a detailed study of the future sensitivity of these observables at FCC-$ee$ and HL-LHC with a particular focus on assumptions of improvements in systematic uncertainties.

We presented an updated SM prediction for the $\tau$ weak magnetic dipole moment at one loop, finding
\begin{equation}\begin{split}
    &\rre\left(\mu_\tau^w\right) = - 2.0757(586)_{\text{scheme}}(160)_{\text{scale}}(6)_{\text{param}}\times 10^{-6}\,,\\[0.5em]
    &\iim\left(\mu_\tau^w\right) = - 0.5997(586)_{\text{scheme}}(71)_{\text{scale}}(3)_{\text{param}}\times 10^{-6}\,,
\end{split}\end{equation}
This result is in good agreement with Ref.~\cite{Bernabeu:1994wh}, and extends it by providing an assessment of the theoretical errors including an estimated uncertainty from explicit electroweak scheme dependence. 

Next, we performed a SMEFT likelihood analysis of the $\tau$ dipole operators $\mathcal{O}_{eW}^{33}$ and $\mathcal{O}_{eB}^{33}$ in the Warsaw basis, finding dominant constraints from the $\tau$ weak and electromagnetic dipole moments, high-mass Drell-Yan tails at the LHC, $Z$ partial decay widths via the double ratio $R_{\tau/\mu}$, and the electron EDM $d_e$, which enters through RG-induced operator mixing. We find that the $\tau$ weak dipole moments play a crucial role in the global fit: $\mu_\tau^w$ provides the most sensitive single constraint in the real plane of $C_{eW}^{33}-C_{eB}^{33}$ parameter space, while the weak electric dipole moment, $d_\tau^w$, is the most important complementary observable to the electron EDM in the imaginary plane, together closing the fit. Drell-Yan processes at the LHC provide important but less sensitive constraints, particularly in the real plane, while $R_{\tau/\mu}$ is not competitive with the weak dipole moments at current experimental precision.

Finally, we assessed the future sensitivity to the $\tau$ weak dipole moments and complementary observables at FCC-$ee$ and HL-LHC. For the prospects of directly probing the $\tau$ weak dipole moments at FCC-$ee$, we find systematic uncertainties will play a critical role given the large statistical improvement at the Tera-Z run of FCC-$ee$. Concretely, a reduction of systematic uncertainties with respect to ALEPH by factors of approximately $140$ and $500$ would be required to probe the SM value of $\re(\mu_\tau^w)$ at $1\sigma$ and $2\sigma$ significance, respectively, setting a clear target for future dedicated experimental sensitivity studies. For the SMEFT projections, even under the conservative assumption that systematic uncertainties are improved by a factor of $10$ with respect to ALEPH, the sensitivity to the $\tau$ dipole operators improves considerably at FCC-$ee$. We find that $\mu_\tau^w$ remains by far the most sensitive constraint in the real-real $C_{eW}^{33}-C_{eB}^{33}$ plane, with important complementary constraints from Drell-Yan processes and $a_\tau$. In the imaginary plane, $d_\tau^w$ and $d_e$ can reach comparable sensitivity and together close the fit completely.

Our results highlight several directions for future work. On the theory side, a full two-loop electroweak computation of $\mu_\tau^w$ would reduce the scheme uncertainty identified in this work, though this is unlikely to be necessary given the expected experimental precision achievable at FCC-$ee$. More urgently, a dedicated computation of the two-loop matching contributions of the $\tau$ dipole operators to the electron EDM would solidify the constraints in the $(\im [C_{eB}^{33}],\im [C_{eW}^{33}])$ plane, where these effects have been shown to be significant. On the experimental side, a dedicated analysis of the systematic uncertainties at FCC-$ee$ would be needed to determine whether the sensitivity targets identified in this work are achievable in practice. Taken together, the $\tau$ weak dipole moments represent a uniquely powerful and experimentally accessible probe of New Physics that will play an increasingly important role in the precision physics program of future colliders.

\section*{Acknowledgements}
The authors acknowledge the financial support from the Slovenian Research and Innovation Agency (grants No.~J1-50219, N1-0407 and research core funding No.~P1-0035).

\appendix

\section{Analytic Results for $F^w_2(M_Z^2)$}~\label{app:analyticRes}
The result for the single photon diagram in Fig.~\ref{fig:dipolefeyndias} is simple enough to be given exactly
\begin{equation}
    F_2^{w,(\gamma)}(M_Z^2) = \frac{\alpha}{4\pi}\frac{g_R + g_L}{2}\frac{1-\beta_\tau^2}{\beta_\tau}\Bigg[i\pi + \log\Big(\frac{1 - \beta_\tau}{1 + \beta_\tau}\Big)\Bigg] \,,
\end{equation}
where
\begin{equation}
    \beta_\tau = \sqrt{1 - \frac{4m_\tau^2}{M_Z^2}} \,,
\end{equation}
is the $\tau$ velocity in the decaying $Z$ rest frame and the right- and left-handed couplings of the $\tau$ to the $Z$ are defined in Eq.~\eqref{eq:gRgL}.

For the remaining diagrams, we present the leading results in $x=m_\tau^2/M_Z^2$, which are precise up to $x\log x \lesssim 1\%$ corrections. For the neutral-current diagrams, we find
\begin{equation}\begin{split}
    F^{w,(Z)}_2(M_Z^2) =\frac{\alpha}{4\pi}x\Bigg[&
    \frac{1}{s_w^3 c_w^3}F^{w,(Z)}_{33} + \frac{1}{s_w c_w^3} F^{w,(Z)}_{13} + \frac{s_w^3}{c_w^3} F^{w,(Z)}_{-33} + \frac{s_w}{c_w^3}F^{w,(Z)}_{-13}\Bigg]\,,
\end{split}\end{equation}
with
\begin{equation}\begin{split}
    F^{w,(Z)}_{33} &= \frac{1}{2} z \text{Li}_2\left(\frac{2}{2-r_- \sqrt{z}}\right)-\frac{1}{2} z \text{Li}_2\left(\frac{2}{r_+ \sqrt{z}}\right)-4 \text{Li}_2(2)+\frac{1}{4} z \log
   ^2\left(\frac{r_- \sqrt{z}}{r_- \sqrt{z}-2}\right) \\[0.5em]
   &\,\,\,\phantom{=}+\frac{1}{2} \sqrt{z-4} \sqrt{z} \log \left(\frac{r_+}{2}\right)+\frac{\pi ^2 z}{12}-\frac{1}{4} z \log
   (z)+\frac{4 \pi ^2}{3}-\frac{11 i \pi }{4}-\frac{21}{8}\,,
\end{split}\end{equation}
\begin{equation}\begin{split}
    F^{w,(Z)}_{13} &=-2 z \text{Li}_2\left(\frac{2}{2-r_- \sqrt{z}}\right)+2 z \text{Li}_2\left(\frac{2}{r_+ \sqrt{z}}\right)+21 \text{Li}_2(2)-z \log ^2\left(\frac{r_-
   \sqrt{z}}{r_- \sqrt{z}-2}\right)\\[0.5em]
    &\,\phantom{=}\,\,-2 \sqrt{z-4} \sqrt{z} \log \left(\frac{r_+}{2}\right)-\frac{\pi ^2 z}{3}+z \log (z)-7 \pi ^2+\frac{29 i \pi }{2}+\frac{59}{4}\,,
\end{split}\end{equation}
\begin{equation}\begin{split}
    F^{w,(Z)}_{-33} &= \,40 \text{Li}_2(2)+34+28 i \pi -\frac{40 \pi ^2}{3}\,,\hspace{7cm}
\end{split}\end{equation}
\begin{equation}\begin{split}
    F^{w,(Z)}_{-13} =& \,-30 \text{Li}_2(2)-\frac{51}{2}-21 i \pi +10 \pi ^2\,,\hspace{6.5cm}
\end{split}\end{equation}
where $z=M_h^2/M_Z^2$, $\text{Li}_2(z)$ is the usual dilogarithm
\begin{equation}
    \text{Li}_2(z) = - \int_0^zdt\,\frac{\log(1 - t)}{t}\,,
\end{equation}
and
\begin{equation}
    r_\pm = \sqrt{z}\pm\sqrt{z-4}\,.
\end{equation}
For the charged-current diagrams, we find
\begin{equation}\begin{split}
    F^{w,(W)}_2(M_Z^2) =\frac{\alpha}{4\pi}x\Bigg[&
    \frac{1}{s_w^3 c_w}F^{w,(W)}_{31} + \frac{c_w}{s_w^3} F^{w,(W)}_{3-1} + \frac{1}{s_wc_w} F^{w,(W)}_{11}\Bigg]\,,
\end{split}\end{equation}
with
\begin{equation}\begin{split}
    F^{w,(W)}_{31} &= \left(3 y^2+4 y+1\right) \text{Li}_2(y+1)-3 y^2 \log (x) \log (y)+3 y^2 \log (x) \log (y+1)\\[0.5em]
    &\,\,\,\phantom{=}-i \left(3 y^2+4 y+1\right) (\pi -i \log (x)) \log \left(\frac{1}{y}+1\right)-4 y \log (x) \log (y)\\[0.5em]
    &\,\,\,\phantom{=}+4 y \log (x) \log (y+1)-\log (x) \log (y)+\log (x) \log (y+1)-\pi ^2 y^2\\[0.5em]
    &\,\,\,\phantom{=}+3 y^2 \log ^2(y)+\frac{3}{2} y^2 \log ^2(y+1)-\frac{1}{2} \left(3 y^2+4 y+1\right) \log ^2\left(\frac{1}{y}+1\right)\\[0.5em]
    &\,\,\,\phantom{=}-3 y^2 \log (y) \log (y+1)+3 i \pi  y^2 \log (y+1)-\frac{4 \pi ^2 y}{3}+3 i \pi  y+3 y\\[0.5em]
    &\,\,\,\phantom{=}+4 y \log ^2(y)+2 y \log ^2(y+1)+\log ^2(y)+\frac{1}{2} \log ^2(y+1)+3 y \log (y)\\[0.5em]
    &\,\,\,\phantom{=}-4 y \log (y) \log (y+1)+4 i \pi  y \log (y+1)+\frac{5 \log (y)}{2}-\log (y) \log (y+1)\\[0.5em]
    &\,\,\,\phantom{=}+i \pi  \log (y+1)-\frac{\pi ^2}{3}+\frac{13}{4}+\frac{5 i \pi }{2}\,,
\end{split}\end{equation}
\begin{equation}\begin{split}
    F^{w,(W)}_{3-1} &= -6 y^2 \text{Li}_2\left(-\frac{2 y}{R_--2 y}\right)-(6 y+1) y \text{Li}_2\left(\frac{2 (y-1)}{R_-}\right)+(6 y+1) y \text{Li}_2\left(\frac{2 (y-1)}{R_--2 y}\right)\\[0.5em]
    &\,\,\,\phantom{=}-y \text{Li}_2\left(-\frac{2 y}{R_--2 y}\right)+6 y^2 \text{Li}_2\left(\frac{2 y}{R_+}\right)+y \text{Li}_2\left(\frac{2 y}{R_+}\right)+6 y^2 \text{Li}_2\left(\frac{y-1}{y}\right)\\[0.5em]
    &\,\,\,\phantom{=}+y \text{Li}_2\left(\frac{y-1}{y}\right)+3 y^2 \log ^2\left(-\frac{R_+}{R_--2 y}\right)+\frac{1}{2} y \log ^2\left(-\frac{R_+}{R_--2 y}\right)\\[0.5em]
    &\,\,\,\phantom{=}-3 y^2 \log ^2\left(\frac{R_-}{R_--2 y}\right)-\frac{1}{2} y \log ^2\left(\frac{R_-}{R_--2 y}\right)-6 \sqrt{1 - 4 y}\, y \log \left(\frac{R_+}{2 y}\right)\\[0.5em]
    &\,\,\,\phantom{=}-2 \sqrt{1 - 4 y} \log \left(\frac{R_+}{2 y}\right)-\pi ^2 y^2-\frac{\pi ^2 y}{6}-6 y-\frac{3}{2}\,,
\end{split}\end{equation}
\begin{equation}\begin{split}
    F^{w,(W)}_{11} &= y \text{Li}_2\left(\frac{2 (y-1)}{R_-}\right)-y \text{Li}_2\left(\frac{2 (y-1)}{R_--2 y}\right)+y \text{Li}_2\left(-\frac{2 y}{R_--2 y}\right)-y \text{Li}_2\left(\frac{2 y}{R_+}\right)\\[0.5em]
    &\,\,\phantom{=}-y \text{Li}_2\left(\frac{y-1}{y}\right)-\frac{1}{2} y \log ^2\left(-\frac{R_+}{R_--2 y}\right)+\frac{1}{2} y \log ^2\left(\frac{R_-}{R_--2 y}\right)\\[0.5em]
    &\,\,\phantom{=}+ \sqrt{1-4 y} \log \left(\frac{R_+}{2 y}\right)+\frac{\pi ^2 y}{6}+1\,,
\end{split}\end{equation}
where $y=M_W^2/M_Z^2$\footnote{Although this ratio is equal to $c_w^2$, we find it slightly preferable to keep separate the mixing angles which arise in coupling constants from mass ratios coming from the propagating intermediate states and kinematics.}
and
\begin{equation}
    R_\pm = -1+2 y \pm \sqrt{1 - 4y}\,.
\end{equation}

\bibliographystyle{JHEP}
\bibliography{draft}

\end{document}